\documentclass[12pt]{article}
\usepackage{blindtext}
\usepackage{amsmath}
\usepackage{amssymb}
\usepackage{wrapfig}
\usepackage{multicol}
\usepackage[margin=1in]{geometry}
\usepackage{graphicx}
\usepackage{caption}
\usepackage{indentfirst}
\usepackage{slashed}

\graphicspath{{Figures/}}
\emergencystretch\maxdimen
\renewenvironment{abstract}
{\par\noindent\textbf{\abstractname:}\ \ignorespaces}
{\par\medskip}

\makeatletter
\renewcommand\maketitle
{\noindent
{\large\bfseries\@title}%
\medskip\par
{\@author}%
\medskip\par
{\@date}%
\bigskip\par\noindent
}

\renewcommand\section{\@startsection {section}{1}{\z@}%
{-1.5ex}%
{0.05ex }%
{\bfseries}}
\makeatother

\begin{document}
\onecolumn
\title{A relativistic statistical field theory on a discretized Minkowski lattice with quantum-like observables}
\author{Brenden McDearmon }
\maketitle

\begin{abstract}
A relativistic statistical field theory is constructed for a fluctuating complex-valued scalar field on a discretized Minkowski lattice. A Hilbert space of observables is then constructed from functionals of the fluctuating complex-valued scalar field with an inner product defined in terms of expectation values of the functionals. A bosonic Fock space is then constructed from the Hilbert space and creation and annihilation operators that act on the Fock space are defined. The creation and annihilation operators are used to define field operators. These field operators have some interesting quantum-like properties. For example, the field operators do not commute in general and, in the particular case of the free field theory, can be shown to satisfy the microcausality condition. 

\end{abstract}

\section*{Introduction}

Quantum field theory has been studied from various perspectives. By one perspective, a classical relativistic field theory is ``quantized'' by replacing the classical field with corresponding field operators that act on some Hilbert space with imposed commutation or anti-commutation relations that depend on whether the field is fermionic or bosonic \cite{peskin2018introduction}\cite{itzykson2012quantum}. In this canonical quantization picture, observables are calculated in terms of expectation values of the field operators with respect to some state in the Hilbert space \cite{peskin2018introduction}\cite{itzykson2012quantum}. For example, the LSZ reduction formula relates expectation values of time ordered products of field operators to scattering amplitudes \cite{peskin2018introduction}\cite{itzykson2012quantum}.

By a second perspective, these same scattering amplitudes can in turn be calculated using the Feynman path integral  \cite{peskin2018introduction}\cite{itzykson2012quantum}\cite{bailin1993introduction}. In the Feynman path integral approach, the expectation values of time ordered products of field operators are calculated by averaging functionals of the classical field over all possible field configurations weighted by a complex factor $e^{iS^m \slash \hbar}$ involving the matter action $S^m$ \cite{peskin2018introduction}\cite{itzykson2012quantum}\cite{bailin1993introduction}. While some techniques exist for evaluating these real time Feynman path integrals, the oscillatory nature of the complex-valued path integral measure can make numerical evaluation of these expectation values challenging \cite{berges2007lattice} \cite{berges2021complex}. To make these calculations more tractable, the real time Feynman path integrals are often ``Wick rotated'' to imaginary time to provide a real-valued path integral measure \cite{bailin1993introduction}.

Wick rotation converts the real time Feynman path integral on Minkowski spacetime into an imaginary time path integral on 4-dimensional Euclidean space \cite{roepstorff2012path}  \cite{glimm2012quantum}. This provides another perspective on quantum field theory whereby expectation values of observables are calculated by averaging the observable over all configurations of the Euclidean classical field \cite{roepstorff2012path}  \cite{glimm2012quantum}. This averaging is performed using a Euclidean path integral having a real-valued probability density proportional to $e^{- \beta S_E^m}$ where $S_E^m$ is the Wick rotated Euclidean matter action and $\beta$ is a real parameter analogous to the thermodynamic $\beta$ (i.e., inverse temperature) of classical statistical mechanics \cite{roepstorff2012path}  \cite{glimm2012quantum}. An advantage of this Euclidean approach to quantum field theory is that it is readily amenable to non-perturbative calculations by discretizing the system with a Euclidean lattice and evaluating the Euclidean path integral numerically \cite{montvay1994quantum}. A disadvantage, however, is that Euclidean space lacks the causal structure of Minkowski spacetime. Despite this disadvantage,  it is sometimes possible to reconstruct relativistic observables from the Euclidean expectation values \cite{glimm2012quantum} \cite{osterwalder1973axioms} \cite{osterwalder1975axioms}. 

It would be desirable to have a formulation of quantum field theory that is both manifestly relativistic, as in the Feynman path integral approach, and has a real-valued probability density, as in the Euclidean path integral approach. Indeed, some authors have suggested that such a reformulation of quantum field theory may be possible. See, e.g., \cite{tilloy2017interacting}, \cite{floerchinger2010quantum}, \cite{hitschfeld2019probability}, and \cite{kent2013path}. If such a perspective of quantum field theory as a relativistic statistical field theory is possible, it may be more amenable to non-perturbative calculations, owing to the real-valued probability density. Indeed the recent article by Giachello and Gradenigo \cite{giachello2024symplectic} provides some numerical calculations for a relativistic field theory using their ``symplectic quantization'' scheme. See, also, \cite{gradenigo2024symplectic2}, \cite{gradenigo2021symplectic1}, and \cite{gradenigo2021symplectic2}. This ``symplectic quantization'' is similar to the ``variational dynamics'' algorithm used in \cite{mcdearmon2023euclidean1} and \cite{mcdearmon2023euclidean2}, the ``variational dynamics'' additionally including an ``action bath'' coupled to the system. Extending this ``variational dynamics'' approach to a relativistic system, this article constructs a relativistic statistical field theory on a discrete lattice and demonstrates that the resulting algebra of observables has some interesting quantum-like properties\footnote{In a companion article, a momentum space formulation is presented \cite{mcdearmon2024rel}.}.

\section*{Constructing a relativistic statistical field theory on a discretized Minkowski lattice}

Let $\mathcal{M}$ be a regular 4-dimensional unit lattice of $N$ finitely many points, and equip $\mathcal{M}$ with the Minkowski metric $\eta = diag(+,-,-,-)$. A complex-valued field, e.g., $\phi$, is a function that assigns a complex number to each point in $\mathcal{M}$. Suppose that the field $\phi$ is allowed to ``fluctuate'' with respect to some parameter $\lambda \in [0, \infty)$. That is,  $\phi(x, \lambda) \in \mathbb{C}$ for all $ (x,\lambda) \in \mathcal{M} \times [0, \infty)$. To prescribe how $\phi$ flucuates with respect to $\lambda$, it is helpful to consider the derivative $\pi^{(\phi)}(x,\lambda) :=\left.\frac{d \phi(x,\lambda+\epsilon)}{d\epsilon}\right|_{\epsilon=0}$. 

Suppose now that these two fields, $\phi$ and $\pi^{(\phi)}$, exchange action between each other and an ``action bath'' as they fluctuate. Let the exchange of action between the fields and the action bath be mediated by fluctuations of a global real and positive scalar, $s(\lambda) \in (0, \infty)$, and its derivative $\pi^{s}(\lambda) :=\left.\frac{d s(\lambda+\epsilon)}{d\epsilon}\right|_{\epsilon=0}$. This collection of degrees of freedom defines a variational phase space $\Gamma := \{\times_{x \in\mathcal{M}} \phi(x),\times_{x \in\mathcal{M}} \pi^{(\phi)}(x), s, \pi^{(s)}\}$. 

An action for this system is a real-valued function on the variational phase space, $S\left(\gamma \left(\lambda\right)\right) \in  \mathbb{R}$ for all $\gamma(\lambda) \in \Gamma$. Let the action be of the form $S := s(\lambda) \left( S^x(\lambda)-S^0 \right)$ where $S^0$ is the action of the system evaluated at $\lambda =0$ and $S^x$ is defined by the following equation:

\small
 \begin{equation}
\begin{aligned}
 S^x (\lambda):= \sum_{x} \left( \frac{\bar{\pi}^{(\phi)}(x, \lambda)\pi^{(\phi)}(x, \lambda)}{ \left(s(\lambda)\right)^2}\right)+ \frac{\left(\pi^{(s)}(\lambda)\right)^2}{2m_s} + S^m[\phi, \bar{\phi}]+ \frac{n_f}{\beta} ln\left(s(\lambda)\right).
\end{aligned}
\end{equation}
\normalsize

\noindent Here, $n_f$ is the number degrees of freedom of the matter field $\phi$, which in this case is 2N, $m_s$ is a coupling constant taken here to be N, $0<\beta<\infty$ is a parameter that is analogous to the thermodynamic $\beta$ (i.e., inverse temperature) of classical statistical mechanics, and $S^m[\phi, \bar{\phi}]$ is a matter action. In particular, the following matter action will be considered throughout the rest of this article:

 \begin{equation}
\begin{aligned}
S^m[\phi, \bar{\phi}] := S^m_{free}[\phi, \bar{\phi}]+S^m_{int} [\phi, \bar{\phi}], \text{ where }
\end{aligned}
\end{equation}
 \begin{equation}
\begin{aligned}
S^m_{free}[\phi, \bar{\phi}]:=  \sum_x \sum_y \bar{\phi}(y, \lambda) D^{-1}_{\alpha}(y-x) \phi(x,\lambda), \text { and}
\end{aligned}
\end{equation}
 \begin{equation}
\begin{aligned}
S^m_{int} [\phi, \bar{\phi}]:= - \sum_x  \kappa_1\bar{\phi}(x, \lambda) \phi(x, \lambda) + \sum_x  \kappa_2 \left(\bar{\phi}(x, \lambda) \phi(x, \lambda) \right)^2 \text{ with  } \kappa_1, \kappa_2 \in \mathbb{R}.
\end{aligned}
\end{equation}
\noindent  In equation (3), $ D^{-1}_{\alpha}$ is some positive definite Hermitian N by N matrix for each $\alpha >0$ that is defined in terms of its inverse matrix $ D_{\alpha}$ where $ D_{\alpha}$ is required to approach a discretized positive-frequency Wightman distribution as $\alpha \to 0^+$. The positive-frequency Wightman distribution is defined by $D_{W}^+(y-x):=\frac{1}{(2\pi)^3} \int \theta(p^0) \delta(p \cdot p -m^2) e^{-ip \cdot (y-x)} d^4p$ where the dot denotes the Minkowski inner product \cite{dutsch2019classical} \cite{scharf2014finite}. The positive-frequency Wightman distribution is a solution of the Klein-Gordon equation (i.e., $\left( \Box_y+m^2 \right) D_{W}^+(y-x)=0$) \cite{dutsch2019classical}. The positive-frequency Wightman distribution is also positive semi-definite and Hermitian \cite{dutsch2019classical}.

The following are three example definitions for $ D_{\alpha}$ from which a corresponding $ D^{-1}_{\alpha}$ can be obtained by matrix inversion as demonstrated in the examples section below:

\begin{equation}
\begin{aligned}
D^{(A)}_{\alpha}(y-x):=\frac{1}{(2\pi)^3} \int \theta_{\alpha}(p^0) \delta_{\alpha}(p \cdot p -m^2) e^{-ip \cdot (y-x)} d^4p,
\end{aligned}
\end{equation}

\begin{equation}
\begin{aligned}
D^{(B)}_{\alpha}(y-x):=\frac{1}{(2\pi)^3} \int \frac{\delta_{\alpha}(p_0 - E)}{2E} e^{-ip \cdot (y-x)} d^4p, \text{and}
\end{aligned}
\end{equation}

\small
\begin{equation}
\begin{aligned}
&D^{(C)}_{\alpha}(y-x):=
\\& - i \,  sgn(x_0-y_0) \, \left( \frac{ \, \delta\left((y-x)\cdot (y-x)\right)}{4\pi} -\frac{m \, \theta \left((y-x)\cdot (y-x)\right)}{8\pi  \sqrt{(y-x)\cdot (y-x)}} J_1\left(m \sqrt{(y-x)\cdot (y-x)}\right) \right)
\\& +PV \bigg\{ \frac{m \, \theta \left((y-x)\cdot (y-x)\right)}{8\pi  \sqrt{(y-x)\cdot (y-x)}}Y_1 \left(m \sqrt{(y-x)\cdot (y-x)}\right) 
\\& \;\;\;\;\;\;\;\;\;\;\;\;\;\;\;\;\;\;\;\;\;\;\;\;\;\;\;\;\;\;\;\;\;\;\;\;\;\;\;\;\;\;\;+ \frac{m \, \theta \left(-(y-x)\cdot (y-x)\right)}{4\pi^2 \sqrt{-(y-x)\cdot (y-x)}} K_1 \left(m\sqrt{-(y-x)\cdot (y-x)}\right) \bigg \}+\alpha \delta(y-x).
\end{aligned}
\end{equation}
\normalsize
\noindent In equations (5) and (6), $\theta_{\alpha} (b) := \frac{1}{2}+tan^{-1}\left(\frac{b}{\alpha}\right)/\pi $ is a smooth approximation to the step function and $ \delta_{\alpha} (b) := \frac{\alpha}{\pi \left(\alpha^2+b^2\right)}$ is a smooth approximation to the Dirac delta function. The definition of equation (6) is motivated by recognizing that $\theta(p^0) \delta(p \cdot p -m^2)$ is equal to $\theta(p^0) \left(\frac{\delta(p_0-E) }{2E}+\frac{\delta(p_0+E) }{2E}\right)$ which is equal to $\frac{\delta(p_0-E) }{2E}$ where $E:=\sqrt{  p_1^2+p_2^2+p_3^2+m^2}$. See, e.g. \cite{scharf2014finite}. Equation (7) is the sum of $\alpha \delta(x-y)$ and the spacetime representation of the positive-frequency Wightman distribution where $J_1$ is the first order Bessel function of the first kind, $Y_1$ is the first order Bessel function of the second kind, $K_1$ is the first order modified Bessel function of the second kind, and $PV \{ \, \}$ refers to the Cauchy principle value. See, e.g.,  \cite{scharf2014finite} but note that the definition used here agrees with that of \cite{dutsch2019classical} but differs from that of \cite{scharf2014finite} by a factor of $-i$.

Returning to the action, $S$ can be treated mathematically as a Hamiltonian function on the variational phase space with the Hamiltonian differential equations (i.e., equations (8) to (11)) providing a symplectic flow on the variational phase space that conserves the total action of the system. See, e.g., \cite{mcdearmon2023euclidean1} and \cite{mcdearmon2023euclidean2}.

\small
 \begin{equation}
\begin{aligned} 
\dot{\pi}^{(\phi)}(x, \lambda)  :=& \left.\frac{d \pi^{(\phi)}(x, \lambda+ \epsilon)}{d \epsilon}\right|_{\epsilon=0} = -\left.\frac{\partial S}{\partial \bar{\phi}(x)}\right|_{\lambda} = - s(\lambda) \left.\frac{\partial S^m}{\partial \bar{\phi}(x)}\right|_{\lambda}
\\ & = - s(\lambda) \left(\sum_y  D^{-1}_{\alpha}(x-y) \phi(y,\lambda) - \kappa_1 \phi(x, \lambda) +   2 \kappa_2 \left(\bar{\phi}(x, \lambda) \phi(x, \lambda) \right) \phi(x, \lambda)\right)
\end{aligned}
 \end{equation}

 \begin{equation} 
\begin{aligned}
\dot{\phi}(x, \lambda) := \left.\frac{d \phi(x, \lambda+ \epsilon)}{d \epsilon} \right|_{\epsilon=0} = \left.\frac{\partial S}{\partial \bar{\pi}^{(\phi)}(x)}\right|_{\lambda} = \frac{\pi^{(\phi)}(x, \lambda)}{s(\lambda)},
\end{aligned}
\end{equation}

 \begin{equation} 
\begin{aligned}
\dot{\pi}^{(s)}(\lambda) := \left.\frac{d \pi^{(s)}(\lambda+ \epsilon)}{d \epsilon} \right|_{\epsilon=0} = -\left.\frac{\partial S}{\partial s}\right|_{\lambda} =   2 \sum_{x} \left( \frac{\bar{\pi}^{(\phi)}(x, \lambda)\pi^{(\phi)}(x, \lambda)}{ \left(s(\lambda)\right)^2}\right)-  \frac{n_f}{\beta }- \left( S^x(\lambda)-S^0 \right), \text{and}
\end{aligned}
 \end{equation}

 \begin{equation} 
\begin{aligned}
\dot{s}(\lambda) := \left.\frac{d s(\lambda+ \epsilon)}{d \epsilon} \right|_{\epsilon=0} = \left.\frac{\partial S}{\partial \pi^{(s)}}\right|_{\lambda} = \frac{\pi^{(s)}(\lambda)}{m_s}.
\end{aligned}
\end{equation}
\normalsize

Given some state of the system, $\gamma(\lambda_0) \in \Gamma$  at $\lambda=\lambda_0$, the set of differential equations defined by equations (8) to (11) specifies the flow of the system from $\gamma(\lambda_0)$ to a new state $\gamma(\lambda_1) \in \Gamma$  at $\lambda=\lambda_1$. The flow generated by these differential equations can be integrated to generate a system trajectory $\gamma(\lambda) \in \Gamma$.

\section*{Observables and expectation values}
Observables for the system are functionals of the matter field, i.e. $\mathcal O \left[\phi \right] \in \mathbb{C}$. The expectation value of an observable can be calculated by averaging the observable along the trajectory of the system using the following equation:

 \begin{equation} 
\begin{aligned}
\left<\mathcal O \left[\phi\right]\right>_{\lambda}:= \lim_{\Lambda \to \infty}\frac{1}{\Lambda} \int_0^{\Lambda} \mathcal O \left[\phi(\lambda)\right] d \lambda.
\end{aligned}
\end{equation}

Assuming that the flow with respect to $\lambda$ is ergodic, one can also calculate expectation values of the observables using an ensemble average.  Because the action $S:= \left( s(\lambda) \left(S^x(\lambda)-S^0 \right)\right)$ is conserved and, by the definition of $S^0$, is equal to zero for all $\lambda$, the ensemble average is calculated by integrating over the variational phase space using the microcanonical ensemble probability measure defined by:

\begin{equation} 
\begin{aligned}
d\Gamma :=  C  \, \delta \left( s \left(S^x-S^0 \right)\right) ds\,d\pi^{(s)}\,\mathcal{D}\left[Im\{\pi^{(\phi)}\}\right]\,\mathcal{D}\left[Re\{\pi^{(\phi)}\}\right]\,\mathcal{D}\left[Im\{\phi\}\right]\,\mathcal{D}\left[Re\{\phi\}\right]. 
\end{aligned}
\end{equation}

\noindent Here, $\mathcal{D}\left[Im\{\phi\}\right]:=\prod_{x} dIm\{\phi)(x)\}$ is the product of integral measures for the imaginary component of the field $\phi$ at each point $x$,  $\mathcal{D}\left[Re\{\phi\}\right]:=\prod_{x} dRe\{\phi(x)\}$ is the product of integral measures for the real component of the field $\phi$ at each point $x$, $\mathcal{D}\left[Im\{\pi^{(\phi)}\}\right]$ and $\mathcal{D}\left[Re\{\pi^{(\phi)}\}\right]$ are defined analogously, each of the integrals are evaluated from $-\infty$ to $\infty$, and $C$ is a constant. It can be shown that the partition function defined by $\mathcal{Z} := \int d \Gamma$ is given by

\begin{equation}
\begin{aligned} 
 \mathcal{Z} =Z_0 \int e^ {- \beta S^m[\phi, \bar{\phi}] }\,\mathcal{D}\left[Im\{\phi\}\right]\,\mathcal{D}\left[Re\{\phi\}\right] \text { where $Z_0$ is a constant.}
\end{aligned}
\end{equation}

To arrive at equation (14) from the definition $\mathcal{Z} := \int d \Gamma$, first make a change of variables $\underline{\pi}^{(\phi)} := \frac{\pi^{(\phi)}(x)}{s}$. This change of variables is equivalent to a re-scaling of $\lambda$. See, e.g., \cite{bond1999nose} for an analogous construction used in classical statistical mechanics. The change of variables changes the integration measure in the partition function by a factor of $s^{n_f}$ yielding:

\begin{equation}
\begin{aligned} 
\mathcal{Z} & := \int d \Gamma
\\& =C \int s^{n_f} \delta \left( s \left(S^x-S^0 \right)\right) ds\,d\pi^{(s)}\,\mathcal{D}\left[Im\{\underline{\pi}^{(\phi)}\}\right]\,\mathcal{D}\left[Re\{\underline{\pi}^{(\phi)}\}\right]\,\mathcal{D}\left[Im\{\phi\}\right]\,\mathcal{D}\left[Re\{\phi\}\right].
\end{aligned}
\end{equation}

Next, integration over the Dirac delta function with respect to $ds$ can be performed using the identity $ \frac{ d }{ds} \delta\left[f(s)\right]= \frac{\delta\left[ f(s-s')\right]}{\left( \frac{df}{ds}|_{(s')}\right)}$, where $s'$ is the isolated zero of $f(s)$ given by the following equation:

\begin{equation}
\begin{aligned} 
s' =  exp  \bigg[  -\frac{\beta}{n_f} \bigg(\sum_{x} \left( \underline{\bar{\pi}}^{(\phi)}(x)\underline{\pi}^{(\phi)}(x)\right)+ \frac{\left(\pi^{(s)}\right)^2}{2m_s} +S^m[\phi, \bar{\phi}]-S^0\bigg)  \bigg].
\end{aligned}
\end{equation}

Performing this integration with respect to $ds$ gives the following expression for $\mathcal{Z}$ where $C'$ is a new constant:
\small
\begin{equation}
\begin{aligned} 
 C' \int exp  \bigg[ - \beta \bigg(\sum_{x} \left( \underline{\bar{\pi}}^{(\phi)}(x)\underline{\pi}^{(\phi)}(x) \right) & + \frac{\left(\pi^{(s)}\right)^2}{2m_s}+S^m[\phi, \bar{\phi}]  \bigg] 
\\ &  d\pi^{(s)}\,\mathcal{D}\left[Im\{\underline{\pi}^{(\phi)}\}\right]\,\mathcal{D}\left[Re\{\underline{\pi}^{(\phi)}\}\right] \,\mathcal{D}\left[Im\{\phi\}\right]\,\mathcal{D}\left[Re\{\phi\}\right].
\end{aligned}
\end{equation}
\normalsize

Because the integrals with respect to $d\pi^{(s)}$, $\mathcal{D}\left[Im\{\underline{\pi}^{(\phi)}\}\right]$, and $\mathcal{D}\left[Re\{\underline{\pi}^{(\phi)}\}\right]]$ are Gaussian, they can be evaluated to provide the desired equation for the partition function: 

\begin{equation}
\begin{aligned} 
 \mathcal{Z} =Z_0 \int e^{  - \beta S^m[\phi, \bar{\phi}] }\,\mathcal{D}\left[Im\{\phi\}\right]\,\mathcal{D}\left[Re\{\phi\}\right].
\end{aligned}
\end{equation}

The partition function is helpful for calculating expectation values because, by the assumption of ergodicity, the expectation value of some observable $\mathcal{O} \left[\phi \right] $ taken with respect to $\lambda$ agrees with its ensemble average. In equations:
\begin{equation} 
\left<\mathcal O \left[\phi\right]\right>_{\lambda}:= \lim_{\Lambda \to \infty}\frac{1}{\Lambda} \int_0^{\Lambda} \mathcal O \left[\phi(\lambda)\right] d \lambda,
\end{equation}

\begin{equation}
\begin{aligned} 
\left<\mathcal O \left[\phi \right]\right>_{\Gamma}:=
 Z_0 \int \mathcal O \left[\phi \right] e^{ - \beta  S^m[\phi, \bar{\phi}] }\mathcal{D}\left[Im\{\phi\}\right]\mathcal{D}\left[Re\{\phi\}\right], \text{and}
\end{aligned}
\end{equation}

\begin{equation}
\begin{aligned} 
 \left<\mathcal O \left[\phi \right]\right>_{\lambda}= \left<\mathcal O \left[\phi \right]\right>_{\Gamma}.
\end{aligned}
\end{equation}

\noindent Because the expectation value with respect to $\lambda$ agrees with the ensemble average, let $\left<\mathcal O \left[\phi \right]\right>$ denote the expectation value determined by either way. 

\section*{A bosonic Fock space of observables}

A bosonic Fock space can be constructed using the set of observables together with an inner product evaluated by taking the expectation value. To see this, define the inner product of observables by $\left<\mathcal O_1 ,\mathcal O_2 \right> := \left<\overline{\mathcal O_1 \left[\phi \right]}\mathcal O_2 \left[\phi \right] \right>$. This inner product has the following properties:

\begin{equation}
\begin{aligned} 
\left<\mathcal O_1,\mathcal O_2 \right>=\overline{\left<\mathcal O_2 ,\mathcal O_1 \right>},
\end{aligned}
\end{equation}

\begin{equation}
\begin{aligned} 
 \left<z \mathcal O_1 ,\mathcal O_2 \right>=\overline{z}\left<\mathcal O_1 ,\mathcal O_2  \right> \text { where } z \in \mathbb{C}, 
 \end{aligned}
\end{equation}

 \begin{equation}
\begin{aligned} 
\left<\mathcal O_1 ,z \mathcal O_2  \right>=z\left<\mathcal O_1 ,\mathcal O_2 \right> \text { where, again, } z \in \mathbb{C},
\end{aligned}
\end{equation}

\begin{equation}
\begin{aligned}  
\left<\mathcal O_1 +\mathcal O_2 , \mathcal O_3  \right>= \left<\mathcal O_1, \mathcal O_3  \right>+\left<\mathcal\mathcal O_2 , \mathcal O_3 \right>, 
 \end{aligned}
\end{equation}

 \begin{equation}
\begin{aligned} 
 \left<\mathcal O_1 , \mathcal O_2 + \mathcal O_3  \right>= \left<\mathcal O_1 , \mathcal O_2\right>+ \left<\mathcal O_1 , \mathcal O_3 \right>, 
\text{and}
\end{aligned}
\end{equation}

\begin{equation}
\begin{aligned} 
\left<\mathcal O_1 ,\mathcal O_1  \right> \geq 0.
\end{aligned}
\end{equation}

Thus, one can define a Hilbert space, $\mathcal{H}_{\mathcal{M}}$, where the elements of the Hilbert space are the observables modulo those observables such that $\left<\mathcal O,\mathcal O \right> = 0$. 

The bosonic Fock space over $\mathcal{M}$ , denoted $\mathcal{F}_{\mathcal{M}}$, can be constructed from the Hilbert space $\mathcal{H}_{\mathcal{M}}$ using the construction provided in \cite{bar2007wave}. Briefly, the inner product on $\mathcal{H}_{\mathcal{M}}$ induces an inner product, denoted $\left<F^{(j)}\big{|}  G^{(j)}  \right >$,  on elements, $F^{(j)}, G^{(j)}$, of the symmetric tensor product space, $\odot^j \mathcal{H}_{\mathcal{M}}$, where $j$ is a positive integer. The inner product is defined by the following equations:

\begin{equation}
\begin{aligned}
\text{for  j $\geq$ 1:} & \left<F_1^{(1)} \odot . . . \odot F_j^{(1)} \big{|} G_1^{(1)}\odot . . . \odot G_j^{(1)}\right > 
\\& := \sum_{Perm\{1, . . .,j\}} \left<F_1^{(1)}, G_{Perm(1)}^{(1)}\right > . . .  \left<F_j^{(1)}, G_{Perm(j)}^{(1)}\right >,\text{ and}
\end{aligned}
\end{equation}

\begin{equation}
\begin{aligned} 
 &\text{ for j = 0:} \left<F^{(0)} | G^{(0)}\right > := \overline{F^{(0)}} G^{(0)} \text{ where }  F^{(0)}, G^{(0)} \in \odot^0 \mathcal{H} := \mathbb{C}.
\end{aligned}
\end{equation}

The bosonic Fock space is defined as the direct sum $\mathcal{F}_{\mathcal{M}}:=\oplus_{j=0}^{\infty} \odot^j \mathcal{H}_{\mathcal{M}}$ and is a Hilbert space with the induced inner product, $\left< F | G \right >:= \sum_{j=0}^{\infty} \left<F^{(j)} |  G^{(j)}\right >$, where the elements, $F, G \in \mathcal{F}_{\mathcal{M}}$, are sequences, $\left( F^{(0)}, F^{(1)}, F^{(2)}, . . .\right)$ and  $\left(G^{(0)}, G^{(1)}, G^{(2)}, . . .\right)$, with each $F^{(j)}, G^{(j)} \in  \odot^j \mathcal{H}_{\mathcal{M}}$. The vacuum vector is given by $\Omega := \left(1, 0, 0, . . . \right)$. 

One can define creation and annihilation operators, $\hat{a}^*(\mathcal O)$ and $\hat{a}(\mathcal O)$ respectively, where $\mathcal O\in \mathcal{H}_{\mathcal{M}}$. The action of $\hat{a}^*(\mathcal O)$ and $\hat{a}(\mathcal O)$ on each $\odot^j \mathcal{H}_{\mathcal{M}} \subset \mathcal{F}_{\mathcal{M}}$ is defined by equations (30) to (32). In equation (31), the tilde denotes that the term $\tilde{F }_i^{(1)}$ is excluded.

\begin{equation}
\begin{aligned} 
 \hat{a}^*(\mathcal O) : \left( F_1^{(1)} \odot . . . \odot F_j^{(1)} \right) \in \odot^j \mathcal{H}_{\mathcal{M}} \rightarrow \left( \mathcal O \odot F_1^{(1)} \odot F_2^{(1)} . . . \odot F_j^{(1)}\right) \in \odot^{j+1} \mathcal{H}_{\mathcal{M}}
\end{aligned}
\end{equation}

\small
\begin{equation}
\begin{aligned} 
\hat{a}(\mathcal O) : \left( F_1^{(1)} \odot . . . \odot F_j^{(1)} \right)  \in \odot^j \mathcal{H}_{\mathcal{M}} \rightarrow \left(\sum_{i=1}^{j} \left<\mathcal O, F_i^{(1)}  \right>F_1^{(1)}  \odot . . . \odot \tilde{F }_i^{(1)} . . . \odot F_j^{(1)}  \right) \in \odot^{j-1} \mathcal{H}_{\mathcal{M}}
\end{aligned}
\end{equation}
\normalsize

\begin{equation}
\begin{aligned} 
 \hat{a}(\mathcal O)\Omega :=0
\end{aligned}
\end{equation}

The creation and annihilation operators are adjoints of each other because $\left< \hat{a}^*(\mathcal O) F | G \right > = \left< F | \hat{a}(\mathcal O)G \right >$. See, e.g., \cite{bar2007wave} at equation 4.9. The creation and annihilation operators satisfy the canonical comutation relations $\left [\hat{a}(\mathcal O_1),\hat{a}(\mathcal O_2) \right]=0$, $\left [\hat{a}^*(\mathcal O_1),\hat{a}^*(\mathcal O_2) \right]=0$, and $\left [\hat{a}(\mathcal O_1),\hat{a}^*(\mathcal O_2) \right]=\left< \mathcal O_1, \mathcal O_2 \right>\hat{\mathcal{I}}_{\mathcal{F}_{\mathcal{M}}}$, where $ \hat{\mathcal{I}}_{\mathcal{F}_{\mathcal{M}}}$ is the identity operator on the Fock space $\mathcal{F}_{\mathcal{M}}$. See, e.g., \cite{bar2007wave} at lemma 4.6.6.

\section*{``Smeared'' field observables and 2-point correlation functions}
A particularly interesting class of observables is the ``smeared'' field observables of the form $\phi(J) := \sum_x \phi(x) J(x)$ where $J \in \mathcal{C} \left(\mathcal{M}, \mathbb{C} \right)$ is a complex function on $\mathcal{M}$. In this case, the inner product $\left< \phi(J), \phi(K) \right>$ is equal to $\sum_x \sum_y \bar{J}(y) \left< \bar{\phi}(y)\phi(x)\right>K(x)$ where $ \left< \bar{\phi}(y)\phi(x)\right>$ is the 2-point correlation function of the statistical field theory. The 2-point correlation function can be calculated by 
\begin{equation}
\begin{aligned} 
\left.\frac{\partial^2 ln \left( \mathcal{Z}\left[J,\bar{J}\right]\right)}{\partial J(y) \partial \bar{J}(x)}\right|_{J,\bar{J}=0} =\left<\bar{\phi}(y)\phi(x)\right>
\end{aligned}
\end{equation}

\noindent where $\mathcal{Z}\left[J,\bar{J}\right]:= \left<e^{\phi(\bar{J})},e^{\phi(\bar{J})}  \right>$ is the moment generating function.

For the particular matter action defined above in equations (2) to (4), the moment generating function is equal to the following:

\begin{equation}
\begin{aligned} 
 &\mathcal{Z}\left[J,\bar{J} \right]=  Z_0 \, exp\left[ - \beta S^m_{int}\left[\frac{\partial}{\partial J(y)}, \frac{\partial}{\partial \bar{J}(x)}\right] \right]  \circ \mathcal{Z}_{free}\left[\bar{J},J \right] , \text{ where }
 \end{aligned}
\end{equation}

\begin{equation}
\begin{aligned} 
 \mathcal{Z}_{free}\left[J,\bar{J} \right] & \propto  \int e^{\phi(\bar{J})+\bar{\phi}(J) - \beta S^m_{free} [\phi, \bar{\phi}] } \,\mathcal{D}\left[Im\{\phi\}\right]\,\mathcal{D}\left[Re\{\phi\}\right], \text{ and }
 \end{aligned}
\end{equation}

\begin{equation}
\begin{aligned} 
S^m_{int}\left[\frac{\partial}{\partial J(y)}, \frac{\partial}{\partial \bar{J}(x)}\right] :=  \left( -\sum_x  \kappa_1 \frac{\partial}{\partial J(x)} \frac{\partial}{\partial \bar{J}(x)} + \sum_x  \kappa_2 \frac{\partial}{\partial J(x)}\frac{\partial}{\partial \bar{J}(x)} \frac{\partial}{\partial J(x)}\frac{\partial}{\partial \bar{J}(x)} \right).
 \end{aligned}
\end{equation}

\noindent To arrive at this result, first recognize that
\begin{equation}
\begin{aligned} 
\mathcal{Z}\left[J,\bar{J}\right] =  \mathcal{Z}_{0} \int \sum_{k=0}^\infty \left(\frac{\left(- \beta S^m_{int}[\phi, \bar{\phi}] \right)^k}{k!}\right)  e^{\phi(\bar{J})+\bar{\phi}(J)} e^{ - \beta S^m_{free}[\phi, \bar{\phi}] }\,\mathcal{D}\left[Im\{\phi\}\right]\,\mathcal{D}\left[Re\{\phi\}\right].
 \end{aligned}
\end{equation}

\noindent Then notice that
\begin{equation}
\begin{aligned} 
S^m_{int}\left[\frac{\partial}{\partial J(y)}, \frac{\partial}{\partial \bar{J}(x)}\right]  \circ e^{\phi(\bar{J})+\bar{\phi}(J)} = S^m_{int} [\phi, \bar{\phi}] e^{\phi(\bar{J})+\bar{\phi}(J)}, \text { and }
 \end{aligned}
\end{equation}

\begin{equation}
\begin{aligned} 
& exp\left[ - \beta S^m_{int}\left[\frac{\partial}{\partial J(y)}, \frac{\partial}{\partial \bar{J}(x)}\right] \right]  \circ e^{\phi(\bar{J})+\bar{\phi}(J)} := \sum_{k=0}^\infty \left(\frac{\left(- \beta S^m_{int}\left[\frac{\partial}{\partial J(y)}, \frac{\partial}{\partial \bar{J}(x)}\right] \right)^k}{k!}\right) e^{\phi(\bar{J})+\bar{\phi}(J)}
\\ &= \sum_{k=0}^\infty \left(\frac{\left(- \beta S^m_{int}[\phi, \bar{\phi}] \right)^k}{k!}\right)  e^{\phi(\bar{J})+\bar{\phi}(J)} .
 \end{aligned}
\end{equation}

\noindent Combining equations (37) and (39) and pulling the operator $exp\left[ - \beta S^m_{int}\left[\frac{\partial}{\partial J(y)}, \frac{\partial}{\partial \bar{J}(x)}\right] \right]$ out of the integral results in equation (34). See, e.g.,\cite{bailin1993introduction} at chapter 5 for a similar construction in the context of the Feynman path integral. 

Notice, in particular, that the 2-point correlation function of the free theory can be evaluated exactly because the Gaussian integral appearing in equation (35) can be evaluated to give

\begin{equation}
\begin{aligned} 
 \mathcal{Z}_{free} \propto e^{\frac{1}{\beta} \sum_x \sum_y \bar{J}(y) D_{\alpha}(y-x) J(x)}, \text { and, thus, }
\end{aligned}
\end{equation}

\begin{equation}
\begin{aligned} 
\left< \bar{\phi}(y)\phi(x)\right>_{free}= \left.\frac{\partial^2 ln \left( \mathcal{Z}_{free}\left[J,\bar{J} \right]\right)}{\partial J(y) \partial \bar{J}(x)}\right|_{J,\bar{J}=0}=\frac{1}{\beta}D_{\alpha}(y-x).
\end{aligned}
\end{equation}

\noindent Equations (33) and (37) can be used to provide a perturbation series for the 2-point correlation function of the interacting theory where the zero-order term is $\left< \bar{\phi}(y)\phi(x)\right>_{free}$.

\section*{Field operators}

Field operators are defined by $\hat{\phi}(J) :=  \hat{a}^*\left(\phi(J)\right)+\hat{a}\left(\phi(J)\right)$. These field operators can be considered to be the ``quantization'' of the smeared field observables $\phi(J)$ because they have many properties that one would like a quantum field theory to satisfy. 

For example, the field operators are self-adjoint because the creation and annihilation operators are adjoints of each other. See, e.g., \cite{bar2007wave}. Also, the field operators do not in general commute. Rather, the commutator of two field operators is given by $\left[\hat{\phi}(J), \hat{\phi}(K) \right]  = 2 i Im \left \{ \left< \phi(J), \phi(K) \right > \right \}  \hat{\mathcal{I}}_{\mathcal{F}_{\mathcal{M}}}$. Moreover, if one defines Hermitian conjugation by $\hat{\phi}(J)^{\dag} :=\hat{\phi}(\bar{J})^*$, the field operators are Hermitian when J is real. As such, the field operators become Hermitian when restricting to the set of observables of the form $\phi(f) := \sum_x \phi(x) f(x)$ where $f \in \mathcal{C} \left(\mathcal{M}, \mathbb{R} \right)$. With this restriction, the commutation relation between the field operators can be simplified to give $\left[\hat{\phi}(f), \hat{\phi}(g) \right]  = \left( 2 i \sum_x \sum_y f(y) g(x) Im \left \{ \left< \bar{\phi}(y)\phi(x) \right>\right \}\right) \hat{\mathcal{I}}_{\mathcal{F}_{\mathcal{M}}}$.

That these field operators $\hat{\phi}(f)$  have additional properties that one would like a quantum field theory to satisfy is further suggested by considering the more simple case of the free field theory (i.e., where $\kappa_1=0$ and $\kappa_2= 0$ in equation (4)) as $\alpha$ approaches $0^+$. In this situation, the 2-point correlation function $\left< \bar{\phi}(y)\phi(x)\right>_{free}$ approaches $\frac{1}{\beta}D_{W}^+(y-x)$.\footnote{Note, that as $\alpha$ approaches $0^+$, $D_{\alpha}^{-1}(y-x)$ becomes singular because $D_{W}^+(y-x)$ has a kernel that at least contains functions of the form $g(x) =\left( \Box+m^2 \right)f(x)$. As such, the matrix inversion used to arrive at equation (40) is not well defined when $\alpha=0$. Thus, ``approaches'' is used here to mean that the 2-point function $\left< \bar{\phi}(y)\phi(x)\right>_{free}$ is approximated by $D_{W}^+(y-x)$ for arbitrarily small but positive $\alpha$.} With this approximation, it can be shown that the field operators satisfy the microcausality condition.

To see this, note that $2 iIm \left \{ \left< \bar{\phi}(y)\phi(x) \right>_{free} \right \}$ is approximated by $\frac{2i}{\beta}Im \left \{D_{W}^+(y-x) \right \}$ as $\alpha$ appraoches $0^+$. This in turn is proportional to the Pauli-Jordan function which is defined by  $D^{PJ}(y-x):=\frac{-i}{(2 \pi)^3} \int sgn(p^0) \delta(p \cdot p-m^2) e^{-ip \cdot (y-x)} d^4p$. See, e.g., \cite{dutsch2019classical}. The Pauli-Jordan function is known to have causal support, i.e. $D^{PJ}(y-x)=0$ if $(y-x) \cdot (y-x) < 0$ \cite{dutsch2019classical}. As such, two field operators $\hat{\phi}(f)$ and $\hat{\phi}(g)$ commute when the supports of $f$ and $g$ are not connected by a causal path in $\mathcal{M}$. Thus, the free theory satisfies the microcausality condition as $\alpha$ approaches $0^+$. 

The free theory also satisfies the time slice condition as $\alpha$ approaches $0^+$. Here, the time slice condition is said to be satisfied when, for two suitable subregions $U_1, U_2 \subset \mathcal{M}$ containing an open neighborhood of a common Cauchy hypersurface $\Sigma_{t_0}$ with $\Sigma_{t_0} \subset U_1 \subset U_2 \subset \mathcal{M}$, the algebra of observables over $U_1$, $\mathcal{A}\left({U_1}\right)$, is isomorphic to the algebra of observables over $U_2$, $\mathcal{A} \left({U_2}\right)$. See, e.g., \cite{bar2007wave} at \S 4.5 for a more precise definition. To see that the time slice condition is satisfied, further assume that the continuum limit $N\rightarrow \infty$ exists and is well-defined, and let $\mathcal{A}\left({U_1}\right)$, respectively $\mathcal{A}\left({U_2}\right)$, be the algebra of observables generated by the field operators $\hat{\phi}(f)$ where the real-valued function $f$ is now a smooth function with compact support contained in $U_1$, respectively $U_2$. The following construction is similar to the more general proof of the time slice condition for the free quantum field theory considered in \cite{bar2007wave}.  

Let $\Sigma_{t_0}:=t^{-1}\left(t_0\right)$ be the Cauchy hypersurface given by a level set of a time function $t: x \in \mathcal{M} \to  \mathbb{R}$. A collection of maps, $h_{\epsilon,t_0} : \mathcal{H}_{\mathcal{M}} \to \mathcal{H}_{\mathcal{M}}$, will be defined that localize $h_{\epsilon,t_0} (f) \in \mathcal{H} _{\mathcal{M}}$ increasingly near the Cauchy hypersurface $\Sigma_{t_0}$ as the parameter $\epsilon$ goes to zero. To construct $h_{\epsilon,t_0}$, first define two functions  $\rho_{\alpha,t_0}^{\pm}(x)$ as follows:
 
 \[\rho_{\epsilon,t_0}^{+}(x):=
\begin{cases}
0  &  t(x) \leq t_0-\epsilon \\
\frac{1}{1+exp \left(\frac{1}{\left(t(x)-t_0\right)+\epsilon}\right)/exp \left(\frac{1}{\epsilon-\left(t(x)-t_0\right)}\right)} & t_0-\epsilon <t(x)<t_0+\epsilon  \\
1 &   t_0+\epsilon \leq t(x)
\end{cases}
\text{, and}\]

 \[\rho_{\epsilon,t_0}^{-}(x):=
\begin{cases}
1  &  t(x) \leq t_0-\epsilon \\
\frac{1}{1+exp \left(\frac{1}{-\left(t(x)-t_0\right)+\epsilon}\right)/exp \left(\frac{1}{\epsilon+\left(t(x)-t_0\right)}\right)} & t_0-\epsilon <t(x)<t_0+\epsilon  \\
0 &   t_0+\epsilon \leq t(x)
\end{cases}
.\]

\noindent Notice that these functions are smooth and form a partition of unity, i.e. $\rho_{\epsilon,t_0}^{+}(x)+\rho_{\epsilon,t_0}^{-}(x)=1$. Further, $\chi_{\epsilon, t_0}(x):= \pm\frac{\partial^2}{\partial x_0^2}\rho_{\epsilon,t_0}^{\pm}(x)$ is a smooth function with support that becomes increasingly more localized near $\Sigma_{t_0}$ as $\epsilon \to 0^+$. Let $D^{adv/ret}(x-y) :=\frac{1}{(2\pi)^3} \int \frac{ e^{-ip \cdot (x-y)}}{ p \cdot p-m^2\mp i sgn(p_0)0^+} d^4p$ be the advanced and retarded propagators for the Klein-Gordon equation satisfying $\left( \Box_x+m^2 \right) D^{adv/ret}(x-y)= -\delta(x-y)$. See, e.g.,  \cite{dutsch2019classical} at \S 1.8. These functions can be used to define the collection of maps $h_{\epsilon,t_0} : \mathcal{H}_{\mathcal{M}} \to \mathcal{H}_{\mathcal{M}}$ as follows:

 \begin{equation}
\begin{aligned} 
 &  h_{\epsilon,t_0}(f) :=
 \\ & \left( \Box_x+m^2 \right) \left(\rho_{\epsilon,t_0}^{+}(x)  \int_{\mathcal{M}} D^{ret}(x-y) f(y) d^4y+\rho_{\epsilon,t_0}^{-}(x)  \int_{\mathcal{M}} D^{adv}(x-y) f(y) d^4y\right)+f(x)
 \\ & = \left(\Box_x \rho_{\epsilon,t_0}^{+}(x)  \right) \int_{\mathcal{M}} D^{ret}(x-y) f(y) d^4y+ \rho_{\epsilon,t_0}^{+}(x) \left( \left(\Box_x +m^2 \right) \int_{\mathcal{M}} D^{ret}(x-y) f(y) d^4y \right)
 \\& \left(\Box_x \rho_{\epsilon,t_0}^{-}(x)  \right) \int_{\mathcal{M}} D^{adv}(x-y) f(y) d^4y+ \rho_{\epsilon,t_0}^{-}(x) \left( \left(\Box_x +m^2 \right) \int_{\mathcal{M}} D^{adv}(x-y) f(y) d^4y \right)
 \\&+f(x)
\\&  =\left(\Box_x \rho_{\epsilon,t_0}^{+}(x)  \right) \int_{\mathcal{M}} D^{ret}(x-y) f(y) d^4y+ \left(\Box_x \rho_{\epsilon,t_0}^{-}(x)  \right) \int_{\mathcal{M}} D^{adv}(x-y) f(y) d^4y 
\\& -\left(\rho_{\epsilon,t_0}^{+}(x)+\rho_{\epsilon,t_0}^{-}(x)\right) f(x) +f(x)
\\& = \chi_{\epsilon,t_0}(x) \int_{\mathcal{M}} D^{ret}(x-y) f(y) d^4y-  \chi_{\epsilon,t_0}(x) \int_{\mathcal{M}} D^{adv}(x-y) f(y) d^4y
\\& = -\chi_{\epsilon,t_0}(x) \int_{\mathcal{M}} D^{PJ}(x-y) f(y) d^4y.
 \end{aligned}
\end{equation}
 
The functions $h_{\epsilon,t_0}(f)$ are real-valued, smooth, and compactly supported within the time slice defined by $U_{\epsilon} := \{x \in \mathcal{M} | t_0-\epsilon <t(x)<t_0+\epsilon \}$ which contains the Cauchy hypersurface $\Sigma_{t_0}$. Moreover, $h_{\epsilon,t_0}(f)-f(x)$ is of the form $\left( \Box_x+m^2 \right) g(x)$ for some smooth, compactly supported function $g(x)$. Thus, $h_{\epsilon,t_0}(f)-f(x) \in Ker(D_{W}^+(y-x))$. As such, the maps $h_{\epsilon,t_0}$ approach isometries as $\alpha$ approaches $0^+$because
 \begin{equation}
\begin{aligned} 
\left<\hat{\phi}\left( h_{\epsilon,t_0}(f)\right),  \hat{\phi}\left(h_{\epsilon,t_0}(g)\right)\right> &=\sum_x \sum_y  h_{\epsilon,t_0}(f)(y) h_{\epsilon,t_0}(g)(x) \left<\bar{\phi}\left(y\right),  \hat{\phi}\left(x\right)\right>_{free}
\\&=\frac{1}{\beta}\sum_x \sum_y  h_{\epsilon,t_0}(f)(y) h_{\epsilon,t_0}(g)(x) D^{+}_W(y-x)
\\&=\frac{1}{\beta}\sum_x \sum_y  f(y) g(x) D^{+}_W(y-x)
\\&=\left< \hat{\phi}\left(f\right),  \hat{\phi}\left(g\right)\right>.
 \end{aligned}
\end{equation}

 Moreover, functions of the form $\left( \Box_x+m^2 \right) g(x)$ for some smooth, compactly supported function $g(x)$ are exactly the kernel of $D^{PJ}(x-y)$ and hence exactly the kernel of the maps $h_{\epsilon,t_0}$. See, e.g., \cite{bar2007wave} at theorem 3.4.7. Accordingly, for any $f,g \in \mathcal{C}^{\infty}_c\left( U_{2},\mathbb{R}\right)$ there exist $ h_{\epsilon,t_0}(f), h_{\epsilon,t_0}(g) \in \mathcal{C}^{\infty}_c\left( U_{\epsilon},\mathbb{R}\right)$ with $\left<\hat{\phi}\left( h_{\epsilon,t_0}(f)\right),  \hat{\phi}\left(h_{\epsilon,t_0}(g)\right)\right> = \left< \hat{\phi}\left(f\right),  \hat{\phi}\left(g\right)\right>$ and $\Sigma_{t_0} \subset U_{\epsilon} \subset U_{2} \subset \mathcal{M}$ for suitably small $\epsilon$. Thus, $\mathcal{A}\left({U_2}\right) \subset \mathcal{A}\left({U_\epsilon}\right)$. Moreover, $ \mathcal{C}^{\infty}_c\left( U_{\epsilon},\mathbb{R}\right) \subset  \mathcal{C}^{\infty}_c\left( U_{2},\mathbb{R}\right)$ because $  U_{\epsilon} \subset   U_{2}$. Thus, $\mathcal{A}\left({U_\epsilon}\right) \subset \mathcal{A}\left({U_2}\right)$. Accordingly, there is an isomorphism of the algebra of observables generated by the field operators over $U_{\epsilon}$ and the algebra of observables generated by the field operators over $U_{2}$ and $\Sigma_{t_0} \subset U_{\epsilon} \subset U_{2} \subset \mathcal{M}$ . Thus, the free field theory satisfies the time slice condition as $\alpha$ approaches $0^+$.

\section*{Examples}

An advantage of the discretized relativistic statistical field theory developed in the previous sections is that it is easy to simulate numerically. In each of the following examples, a regular Minkowski lattice of $N=9 \times 9 \times 9 \times 9$ points having a lattice spacing of $\pm 1$ was used. The complex scalar field $\phi$ was initialized with a value of 0 at each point. The variational conjugate field $\pi^{(\phi)}$ was initialized at each point with a value randomly selected from a uniform distribution of complex numbers having a magnitude less than or equal to 1.75 so that $\frac{\sum \pi^{(\phi)}\bar{\pi}^{(\phi)}}{N}$ was about equal to 1. For every example, $s$ was initialized at 1,  $\pi^{(s)}$ was initialized at 0, $\beta$ was set equal to 1, and $m_s$ was set equal to $N$.

Once an initial configuration was established, each example was numerically integrated with respect to $\lambda$ using the explicit leap-frog algorithm provided by Bond, Leimkuhler, and Laird \cite{bond1999nose} with a step size of $\Delta \lambda=0.01$. Each system was ``equilibrated" by stepping the system forward for 250,000 steps. After this ``equilibration'' period, each system was further numerically integrated with respect to $\lambda$ for an additional 1,000,000 steps during which data was collected.

The following examples serve only to demonstrate proof-of-concept implementation of non-perturbative calculations using the discretized relativistic statistical field theory framework. The examples are not intended to be quantitatively accurate. Quantitative accuracy of these examples may be limited by, for example, the small lattice size of the simulated systems, the number of steps simulated, and the particular discretizations employed.

\subsection*{\small \;\;Example 1:\;}

The system of example 1 was simulated using the matter action $S^m[\phi, \bar{\phi}] := S^m_{free}[\phi, \bar{\phi}]+S^m_{int} [\phi, \bar{\phi}]$ where $S^m_{free}[\phi, \bar{\phi}]=  \sum_x \sum_y \bar{\phi}(y, \lambda) \left(D^{(A)}_{\alpha}\right)^{-1}(y-x) \phi(x,\lambda)$ and $S^m_{int} [\phi, \bar{\phi}]:= - \sum_x  \kappa_1\bar{\phi}(x, \lambda) \phi(x, \lambda) + \sum_x  \kappa_2 \left(\bar{\phi}(x, \lambda) \phi(x, \lambda) \right)^2$. Here, $\kappa_1$ was equal to 0, 1, or 2, $\kappa_2$ was equal to 0 or 1, and $\alpha$ was equal to $\frac{1}{9}$. Recall, the matrix $D^{(A)}_{\alpha}(y-x)$ is defined as $\frac{1}{(2\pi)^3} \int \theta_{\alpha}(p^0) \delta_{\alpha}(p \cdot p -m^2) e^{-ip \cdot (y-x)} d^4p$. Rather than attempting to evaluate the Fourier integrals exactly, $D^{(A)}_{\alpha}(y-x)$ was evaluated numerically with each momentum space integral replaced by a summation ranging from $-\frac{N^{\frac{1}{4}}-1}{2}$ to $\frac{N^{\frac{1}{4}}-1}{2}$. After numerically calculating $D^{(A)}_{\alpha}(y-x)$, the positive definiteness of this matrix was confirmed by Cholesky decomposition. The $D^{(A)}_{\alpha}(y-x)$ matrix was then numerically inverted to obtain $ \left(D^{(A)}_{\alpha}\right)^{-1}(y-x)$. The system was then simulated as discussed above. 

\begin{figure*}[!t]
\centering
\includegraphics[width=0.95\textwidth]{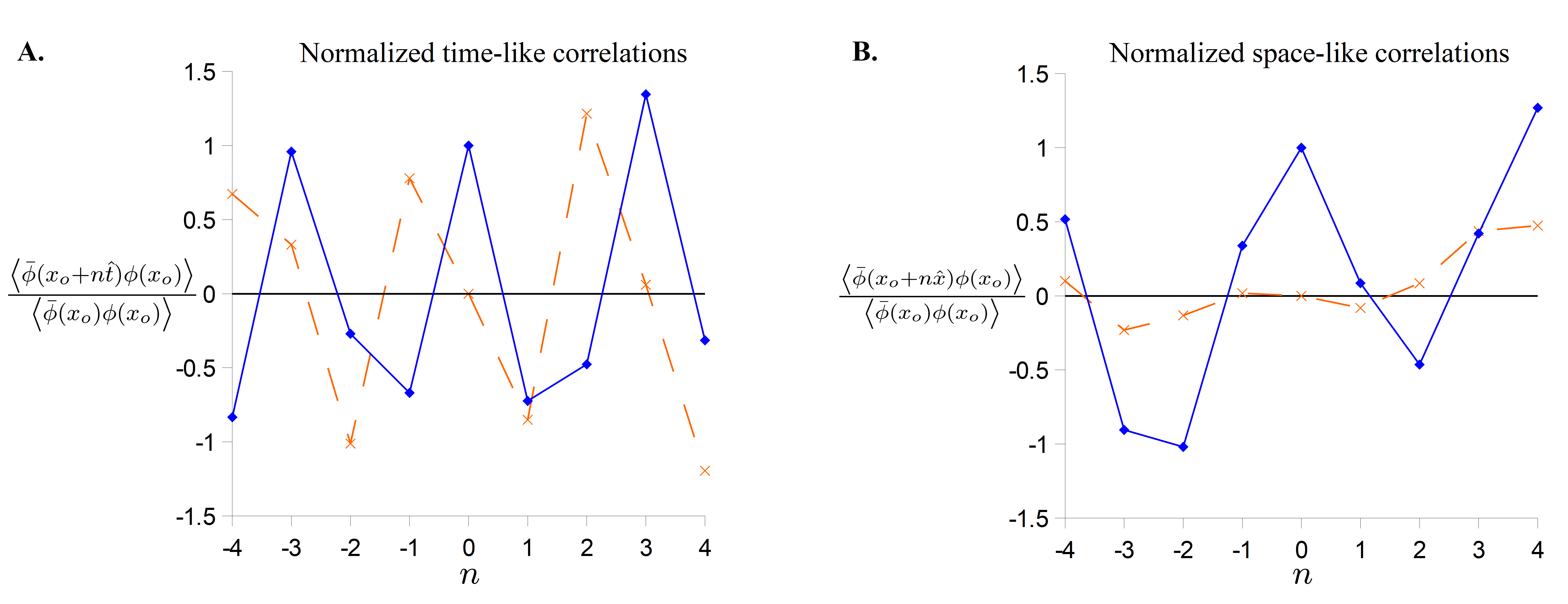}
\caption{Real (\includegraphics[scale=0.1]{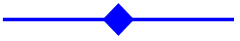}) and imaginary (\includegraphics[scale=0.1]{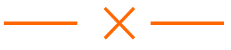}) portions of the time-like (Figure 1A) and space-like (Figure 1B) normalized 2-point correlation function based at the center point ($x_o$) of the Minkowski lattice for the free field theory.}
\end{figure*}

Figure 1A depicts the normalized 2-point correlation function for the free field theory (i.e., $\kappa_1=0$ and $\kappa_2=0$) evaluated along a time-like line passing through the center point of the Minkowski lattice, $x_o$. This time-like line cut of the normalized 2-point correlation function is defined by $\frac{\left<\bar{\phi}(x_o+n \hat{t})\phi(x_o)\right>}{\left<\bar{\phi}(x_o)\phi(x_o)\right>}$ where $\hat{t}$ is a unit lattice vector in the positive time-like direction and $n$ is an integer.  Figure 1B depicts the normalized 2-point correlation function for the free field theory evaluated along a space-like line passing through the center point of the Minkowski lattice. This space-like line cut of the normalized 2-point correlation function is defined by $\frac{\left<\bar{\phi}(x_o+n \hat{x})\phi(x_o)\right>}{\left<\bar{\phi}(x_o)\phi(x_o)\right>}$ where $\hat{x}$ is a unit lattice vector in a positive space-like direction. Both figures 1A and 1B show that the 2-point correlation function of this system has an oscillatory nature in both the time-like and space-like directions. Interestingly, the self correlation, $\left<\bar{\phi}(x_o)\phi(x_o)\right>$, was not a maximum value. Differences of the observed 2-point correlation function from the positive-frequency Wightman distribution $D_{W}^+(y-x)$ may, for example, be due to how the matrix $D^{(A)}_{\alpha}(y-x)$ was regularized at singular points, the value of $\alpha$ used, the discretization of the Fourier integrals, and the limited number of numerical integration steps that were simulated.

\begin{figure*}[!b]
\centering
\includegraphics[width=0.95\textwidth]{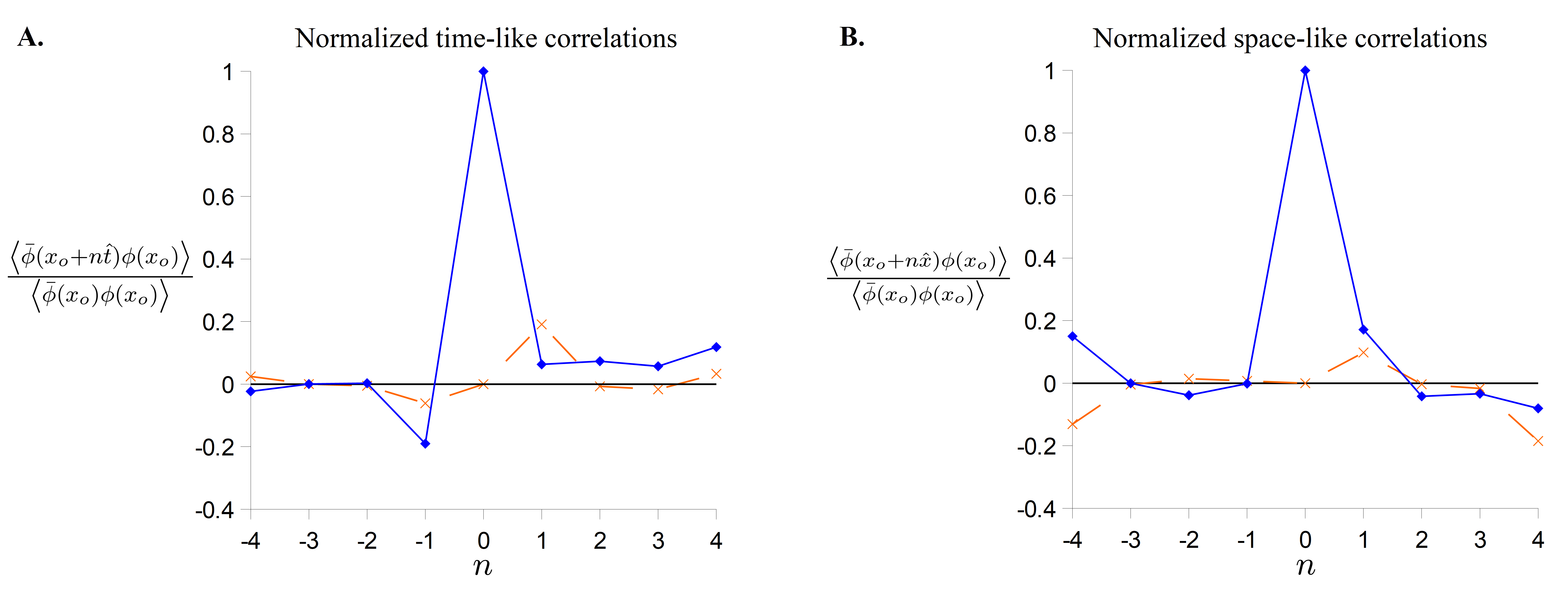}
\caption{Real (\includegraphics[scale=0.1]{Re}) and imaginary (\includegraphics[scale=0.1]{Im}) portions of the time-like (Figure 2A) and space-like (Figure 2B) normalized 2-point correlation function based at the center point ($x_o$) of the Minkowski lattice for the interacting system with $\kappa_1=0$ and $\kappa_2=1$.}
\end{figure*}

Figure 2A depicts the normalized 2-point correlation function for the system with $\kappa_1=0$ and $\kappa_2=1$ evaluated along the time-like line passing through the center point of the Minkowski lattice. Figure 2B depicts the normalized 2-point correlation function for the system with $\kappa_1=0$ and $\kappa_2=1$ evaluated along the space-like line passing through the center point of the Minkowski lattice. Here, unlike in figures 1A and 1B, the self correlation, $\left<\bar{\phi}(x_o)\phi(x_o)\right>$, was maximal with correlations between $x_o$ and other points present but diminished. 

\begin{figure*}[t!]
\centering
\includegraphics[width=0.95\textwidth]{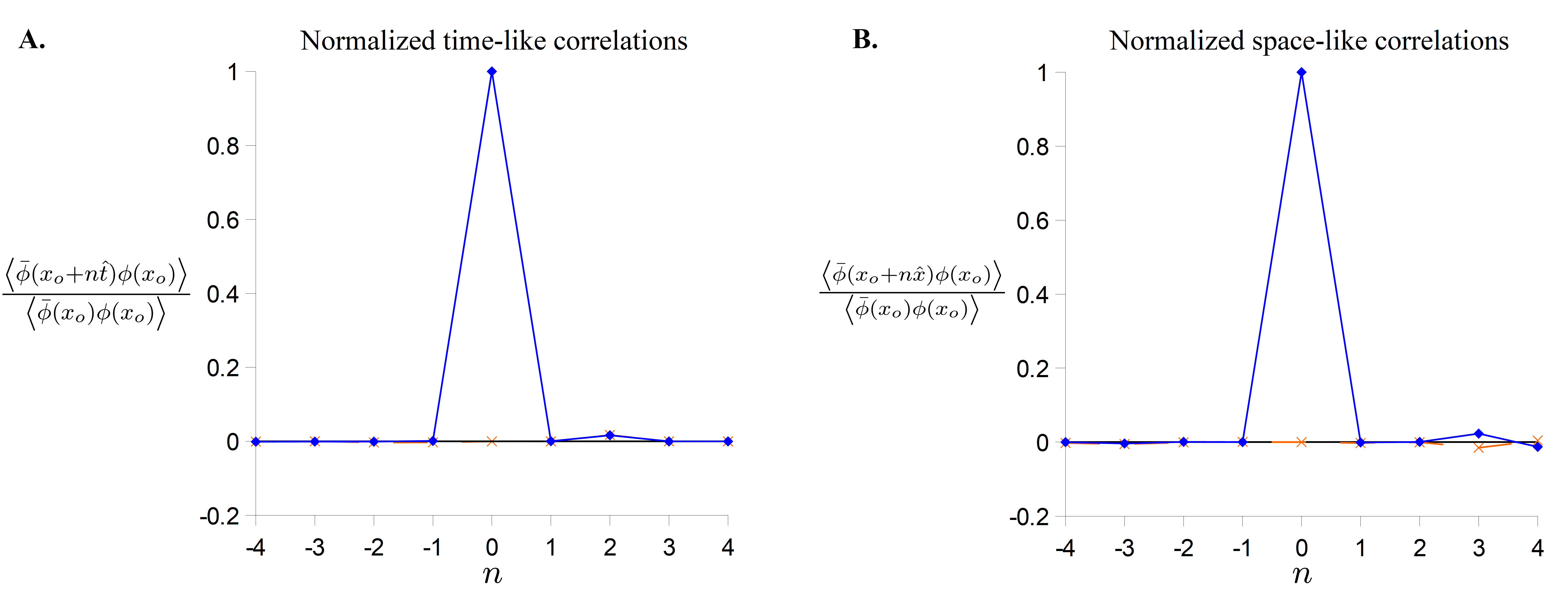}
\caption{Real (\includegraphics[scale=0.1]{Re}) and imaginary (\includegraphics[scale=0.1]{Im}) portions of the time-like (Figure 3A) and space-like (Figure 3B) normalized 2-point correlation function based at the center point ($x_o$) of the Minkowski lattice for the interacting system with $\kappa_1=1$ and $\kappa_2=1$.}
\end{figure*}

Figure 3A depicts the normalized 2-point correlation function for the system with $\kappa_1=1$ and $\kappa_2=1$ evaluated along the time-like line passing through the center point of the Minkowski lattice.  Figure 3B depicts the normalized 2-point correlation function for the system with $\kappa_1=1$ and $\kappa_2=1$ evaluated along the space-like line passing through the center point of the Minkowski lattice. As seen in figures 3A and 3B, this system had much less correlation between field values at different points as compared to the previous systems.

\begin{figure*}[b!]
\centering
\includegraphics[width=0.95\textwidth]{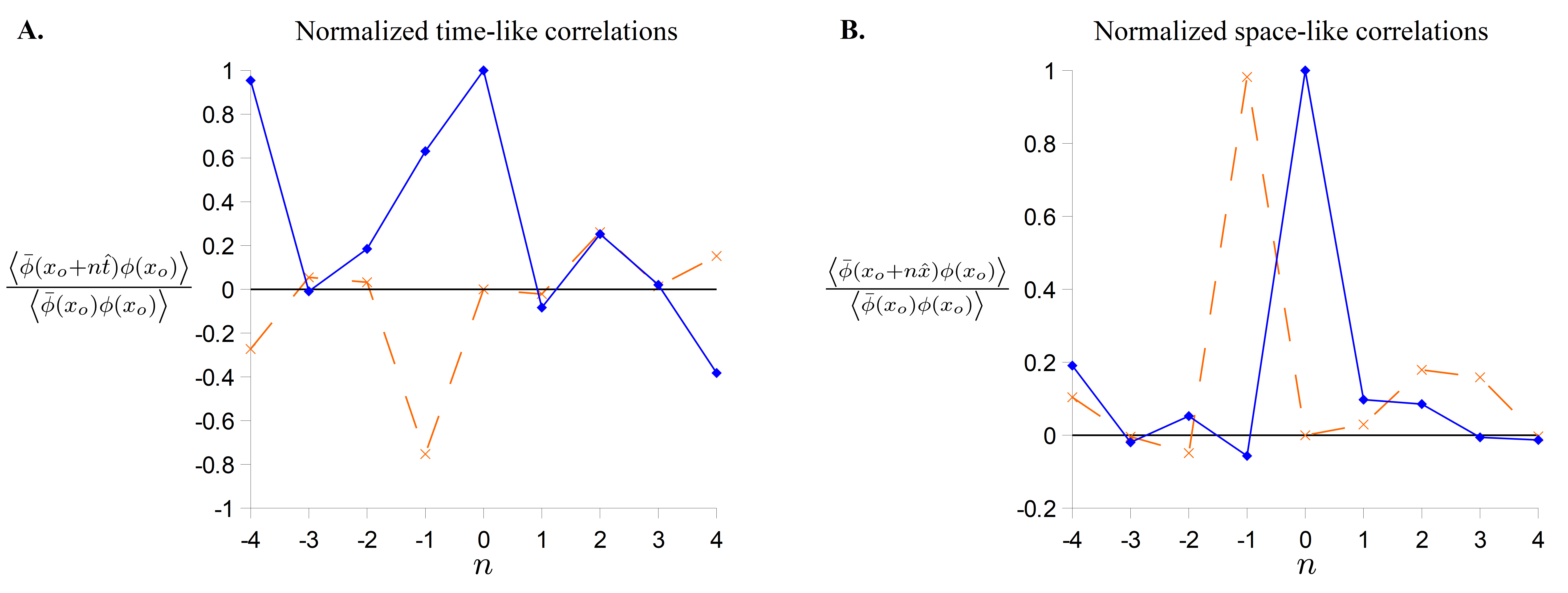}
\caption{Real (\includegraphics[scale=0.1]{Re}) and imaginary (\includegraphics[scale=0.1]{Im}) portions of the time-like (Figure 4A) and space-like (Figure 4B) normalized 2-point correlation function based at the center point ($x_o$) of the Minkowski lattice for the interacting system with $\kappa_1=2$ and $\kappa_2=1$.}
\end{figure*}

Figure 4A depicts the normalized 2-point correlation function for the system with $\kappa_1=2$ and $\kappa_2=1$ evaluated along the time-like line passing through the center point of the Minkowski lattice. Figure 4B depicts the normalized 2-point correlation function for the system with $\kappa_1=2$ and $\kappa_2=1$ evaluated along the space-like line passing through the center point of the Minkowski lattice. This system has an interesting 2-point correlation function that has a different form along the time-like and space-like directions. Comparing Figures 1-4, the transition from a correlated state in figures 1A and 1B to intermediate, less correlated states in figures 2A, 2B, 3A, and 3B, followed further by the emergence of a differently correlated state in figures 4A and 4B suggest that the system may undergo a phase transition along a trajectory in the $(\kappa_1, \kappa_2)$ parameter space.

\subsection*{\small \;\;Example 2:\;}

The system of example 2 was simulated using the matter action $S^m[\phi, \bar{\phi}] := S^m_{free}[\phi, \bar{\phi}]+S^m_{int} [\phi, \bar{\phi}]$ where $S^m_{free}[\phi, \bar{\phi}]$ is equal to $\sum_x \sum_y \bar{\phi}(y, \lambda) \left(D^{(B)}_{\alpha}\right)^{-1}(y-x) \phi(x,\lambda)$ and $S^m_{int} [\phi, \bar{\phi}]$ is equal to $- \sum_x  \kappa_1\bar{\phi}(x, \lambda) \phi(x, \lambda) + \sum_x  \kappa_2 \left(\bar{\phi}(x, \lambda) \phi(x, \lambda) \right)^2$. Here, $\kappa_1$ was equal to 0, 1, or 2, $\kappa_2$ was equal to 0 or 1, and $\alpha$ was equal to $\frac{1}{9}$. Recall, the matrix $D^{(B)}_{\alpha}(y-x)$ is defined as $\frac{1}{(2\pi)^3} \int \frac{\delta_{\alpha}(p_0 - E)}{2E} e^{-ip \cdot (y-x)} d^4p$. Rather than attempting to evaluate the Fourier integrals exactly, $D^{(B)}_{\alpha}(y-x)$ was evaluated numerically with each momentum space integral replaced by a summation ranging from $-\frac{N^{\frac{1}{4}}-1}{2}$ to $\frac{N^{\frac{1}{4}}-1}{2}$. After numerically calculating $D^{(B)}_{\alpha}(y-x)$, the  positive definiteness of this matrix was confirmed by Cholesky decomposition. The  $D^{(B)}_{\alpha}(y-x)$ matrix was the numerically inverted to obtain $ \left(D^{(B)}_{\alpha}\right)^{-1}(y-x)$. The system was then simulated as discussed above.

\begin{figure*}[b!]
\centering
\includegraphics[width=0.95\textwidth]{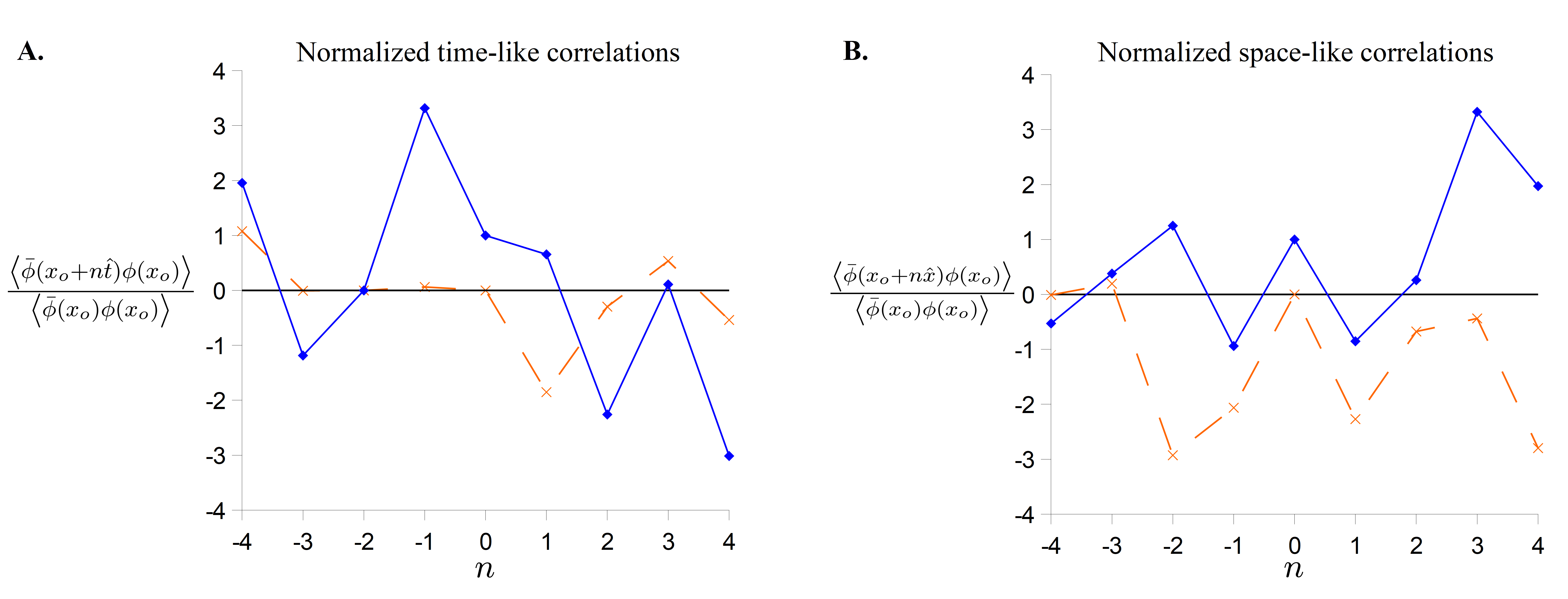}
\caption{Real (\includegraphics[scale=0.1]{Re}) and imaginary (\includegraphics[scale=0.1]{Im}) portions of the time-like (Figure 5A) and space-like (Figure 5B) normalized 2-point correlation function based at the center point ($x_o$) of the Minkowski lattice for the free field theory.}
\end{figure*}

Figure 5A depicts the normalized 2-point correlation function for the free field theory (i.e., $\kappa_1=0$ and $\kappa_2=0$) evaluated along the time-like line passing through the center point of the Minkowski lattice. Figure 5B depicts the normalized 2-point correlation function for the free field theory evaluated along the space-like line passing through the center point of the Minkowski lattice. Both figures 5A and 5B show that the 2-point correlation function of this system has an oscillatory nature in the time-like and space-like directions. Similarly to the free field theory in example 1, the self correlation, $\left<\bar{\phi}(x_o)\phi(x_o)\right>$, was not a maximum value. Again, differences of the observed 2-point correlation function from the positive-frequency Wightman distribution $D_{W}^+(y-x)$ may, for example, be due to how the matrix $D^{(B)}_{\alpha}(y-x)$ was regularized at singular points, the value of $\alpha$ used, the discretization of the Fourier integrals, and the limited number of numerical integration steps that were simulated.

\begin{figure*}[t!]
\centering
\includegraphics[width=0.95\textwidth]{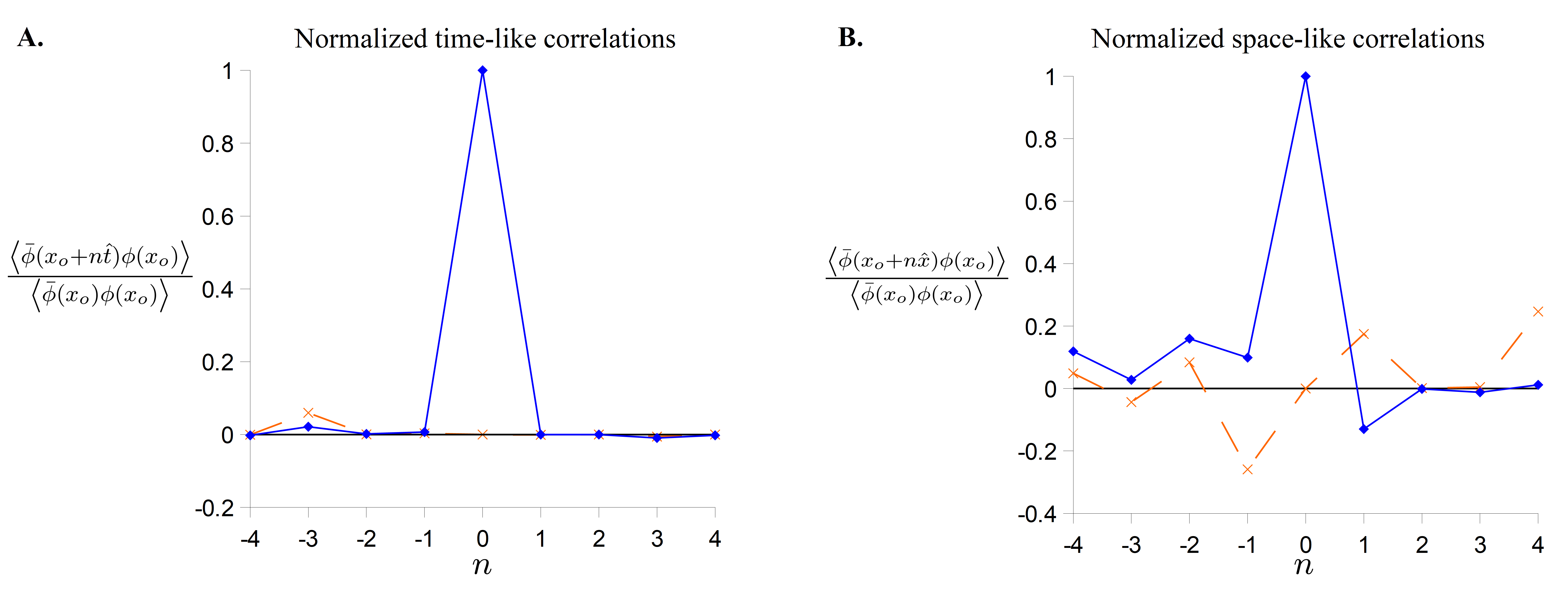}
\caption{Real (\includegraphics[scale=0.1]{Re}) and imaginary (\includegraphics[scale=0.1]{Im}) portions of the time-like (Figure 6A) and space-like (Figure 6B) normalized 2-point correlation function based at the center point ($x_o$) of the Minkowski lattice for the interacting system with $\kappa_1=0$ and $\kappa_2=1$.}
\end{figure*}

\begin{figure*}[b!]
\centering
\includegraphics[width=0.95\textwidth]{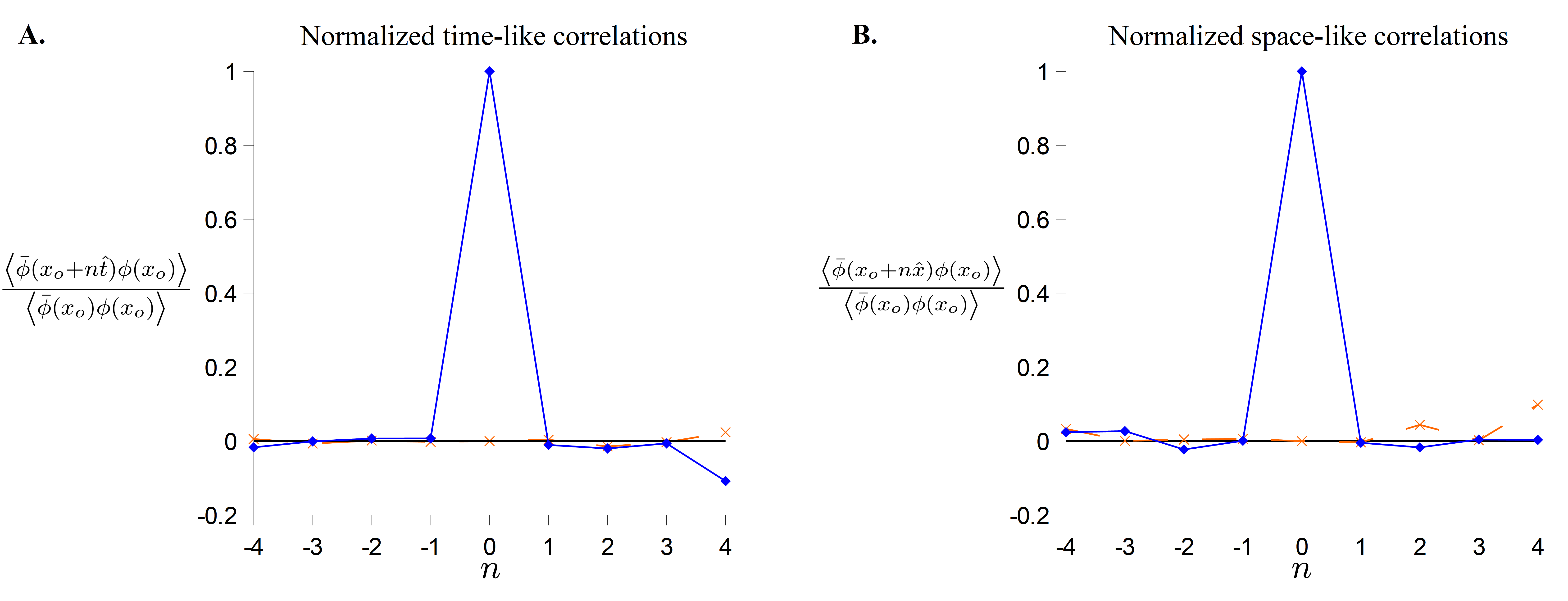}
\caption{Real (\includegraphics[scale=0.1]{Re}) and imaginary (\includegraphics[scale=0.1]{Im}) portions of the time-like (Figure 7A) and space-like (Figure 7B) normalized 2-point correlation function based at the center point ($x_o$) of the Minkowski lattice for the interacting system with $\kappa_1=1$ and $\kappa_2=1$.}
\end{figure*}

Figure 6A depicts the normalized 2-point correlation function for the system with $\kappa_1=0$ and $\kappa_2=1$ evaluated along the time-like line passing through the center point of the Minkowski lattice. Figure 6B depicts the normalized 2-point correlation function for the system with $\kappa_1=0$ and $\kappa_2=1$ evaluated along the space-like line passing through the center point of the Minkowski lattice.

Figure 7A depicts the normalized 2-point correlation function for the system with $\kappa_1=1$ and $\kappa_2=1$ evaluated along the time-like line passing through the center point of the Minkowski lattice. Figure 7B depicts the normalized 2-point correlation function for the system with $\kappa_1=1$ and $\kappa_2=1$ evaluated along the space-like line passing through the center point of the Minkowski lattice.

Figure 8A depicts the normalized 2-point correlation function for the system with $\kappa_1=2$ and $\kappa_2=1$ evaluated along the time-like line passing through the center point of the Minkowski lattice. Figure 8B depicts the normalized 2-point correlation function for the system with $\kappa_1=2$ and $\kappa_2=1$ evaluated along the space-like line passing through the center point of the Minkowski lattice. Similarly to example 1, the transition from a correlated state in figure 5 to intermediate, less correlated states in figures 6 and 7, followed further by the emergence of a differently correlated state in figure 8 suggests that the system may undergo a phase transition along a trajectory in the $(\kappa_1, \kappa_2)$ parameter space.

\begin{figure*}[t!]
\centering
\includegraphics[width=0.95\textwidth]{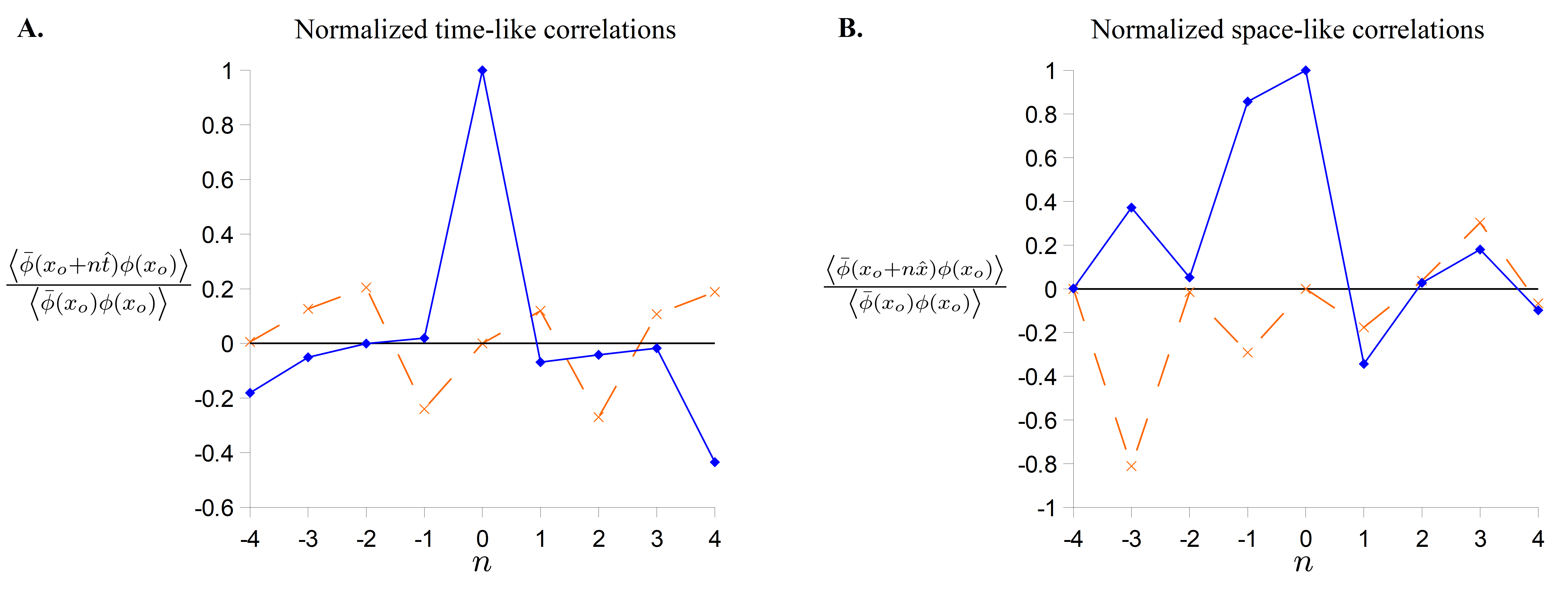}
\caption{Real (\includegraphics[scale=0.1]{Re}) and imaginary (\includegraphics[scale=0.1]{Im}) portions of the time-like (Figure 8A) and space-like (Figure 8B) normalized 2-point correlation function based at the center point ($x_o$) of the Minkowski lattice for the interacting system with $\kappa_1=2$ and $\kappa_2=1$.}
\end{figure*}

\subsection*{\small \;\;Example 3:\;}

The system of example 3 was simulated using the matter action $S^m[\phi, \bar{\phi}] := S^m_{free}[\phi, \bar{\phi}]+S^m_{int} [\phi, \bar{\phi}]$ where $S^m_{free}[\phi, \bar{\phi}]$ is equal to $\sum_x \sum_y \bar{\phi}(y, \lambda) \left(D^{(C)}_{\alpha}\right)^{-1}(y-x) \phi(x,\lambda)$ and $S^m_{int} [\phi, \bar{\phi}]$ is equal to $- \sum_x  \kappa_1\bar{\phi}(x, \lambda) \phi(x, \lambda) + \sum_x  \kappa_2 \left(\bar{\phi}(x, \lambda) \phi(x, \lambda) \right)^2$. Here, $\kappa_1$ was equal to 0, 1, or 2, $\kappa_2$ was equal to 0 or 1, and $\alpha$ was equal to $\frac{1}{9}$. Recall, the matrix $D^{(C)}_{\alpha}(y-x)$ is defined as follows:

\small
\begin{equation}
\begin{aligned}
&D^{(C)}_{\alpha}(y-x):=
\\& - i \,  sgn(x_0-y_0) \, \left( \frac{ \, \delta\left((y-x)\cdot (y-x)\right)}{4\pi} -\frac{m \, \theta \left((y-x)\cdot (y-x)\right)}{8\pi  \sqrt{(y-x)\cdot (y-x)}} J_1\left(m \sqrt{(y-x)\cdot (y-x)}\right) \right)
\\& +PV \bigg\{ \frac{m \, \theta \left((y-x)\cdot (y-x)\right)}{8\pi  \sqrt{(y-x)\cdot (y-x)}}Y_1 \left(m \sqrt{(y-x)\cdot (y-x)}\right) 
\\& \;\;\;\;\;\;\;\;\;\;\;\;\;\;\;\;\;\;\;\;\;\;\;\;\;\;\;\;\;\;\;\;\;\;\;\;\;\;\;\;\;\;\;+ \frac{m \, \theta \left(-(y-x)\cdot (y-x)\right)}{4\pi^2 \sqrt{-(y-x)\cdot (y-x)}} K_1 \left(m\sqrt{-(y-x)\cdot (y-x)}\right) \bigg \}+\alpha \delta(y-x).
\end{aligned}
\end{equation}
\normalsize

The Cauchy principle value was calculated for otherwise singular combinations of $(y,x)$ by averaging the results obtained when displacing the point $y$ by a small real number, here $\frac{1}{36}$, in each direction perpendicular to the singularity. For example, $D^{(C)}_{\alpha}(y-x)$ was evaluated  at $y=x$ by averaging the values obtained at each of the 8 points a distance of $ \pm \frac{1}{36}$ in the each of the 4 coordinate directions. Similarly, the distribution $D^{(C)}_{\alpha}(y-x)$ was evaluated  at $(y-x)\cdot (y-x)=0$ and $x \neq y$ by averaging the values obtained at each of the 2 points at a distance having a magnitude of $ \frac{1}{36}$ in the direction perpendicular to the tangent plane at $y$ of the light-cone containing $x$ and $y$. After calculating $D^{(C)}_{\alpha}(y-x)$, the positive definiteness of this matrix was confirmed by Cholesky decomposition. The matrix $D^{(C)}_{\alpha}(y-x)$ was then numerically inverted to obtain $ \left(D^{(C)}_{\alpha}\right)^{-1}(y-x)$. The system was then simulated as discussed above.

\begin{figure*}[b!]
\centering
\includegraphics[width=0.95\textwidth]{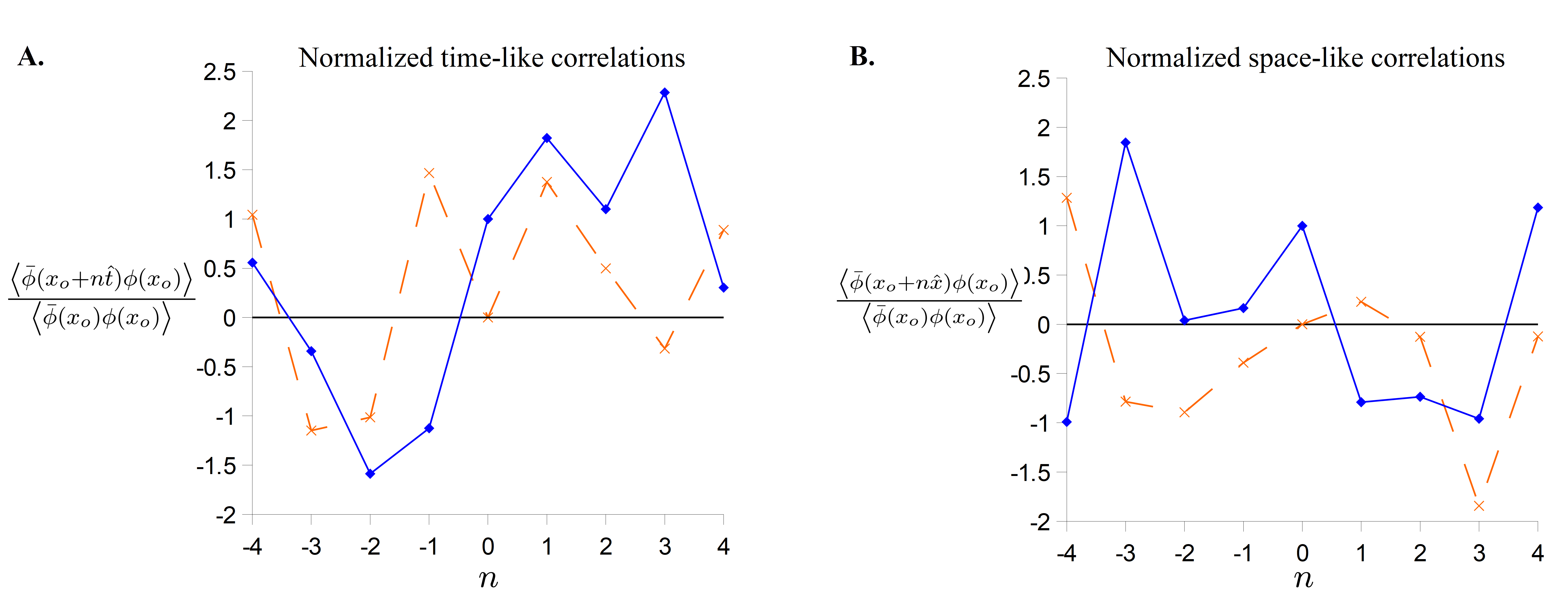}
\caption{Real (\includegraphics[scale=0.1]{Re}) and imaginary (\includegraphics[scale=0.1]{Im}) portions of the time-like (Figure 9A) and space-like (Figure 9B) normalized 2-point correlation function based at the center point ($x_o$) of the Minkowski lattice for the free field theory.}
\end{figure*}

\begin{figure*}[t!]
\centering
\includegraphics[width=0.95\textwidth]{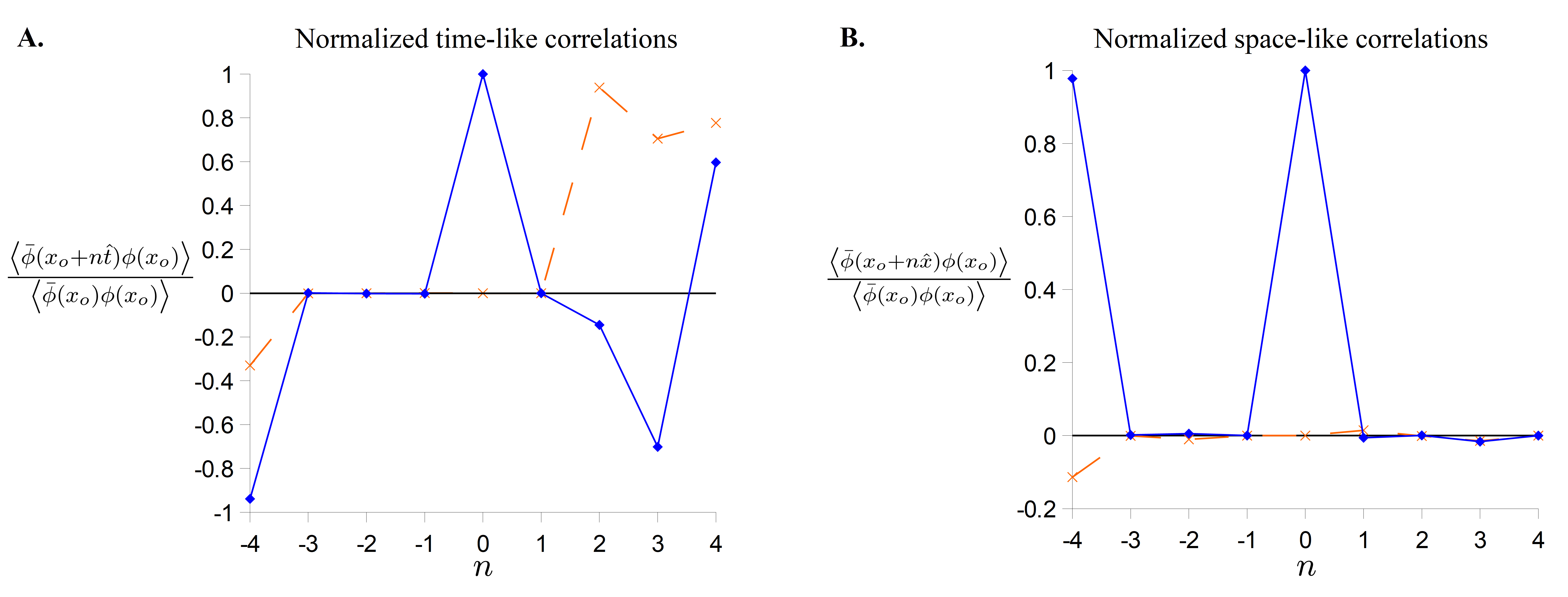}
\caption{Real (\includegraphics[scale=0.1]{Re}) and imaginary (\includegraphics[scale=0.1]{Im}) portions of the time-like (Figure 10A) and space-like (Figure 10B) normalized 2-point correlation function based at the center point ($x_o$) of the Minkowski lattice for the interacting system with $\kappa_1=0$ and $\kappa_2=1$.}
\end{figure*}

Figure 9A depicts the normalized 2-point correlation function for the free field theory (i.e., $\kappa_1=0$ and $\kappa_2=0$) evaluated along the time-like line passing through the center point of the Minkowski lattice.  Figure 9B depicts the normalized 2-point correlation function for the free field theory evaluated along the space-like line passing through the center point of the Minkowski lattice. The features of the 2-point correlation function are qualitatively similar to those of the previous free field examples. The fact that the Fourier integrals of this example were not discretized suggest that differences of the observed 2-point correlation function from the positive-frequency Wightman distribution $D_{W}^+(y-x)$ may be due to how the matrix $D^{(B)}_{\alpha}(y-x)$ was regularized at singular points, the value of $\alpha$ used, and the limited number of numerical integration steps that were simulated.

\begin{figure*}[b!]
\centering
\includegraphics[width=0.95\textwidth]{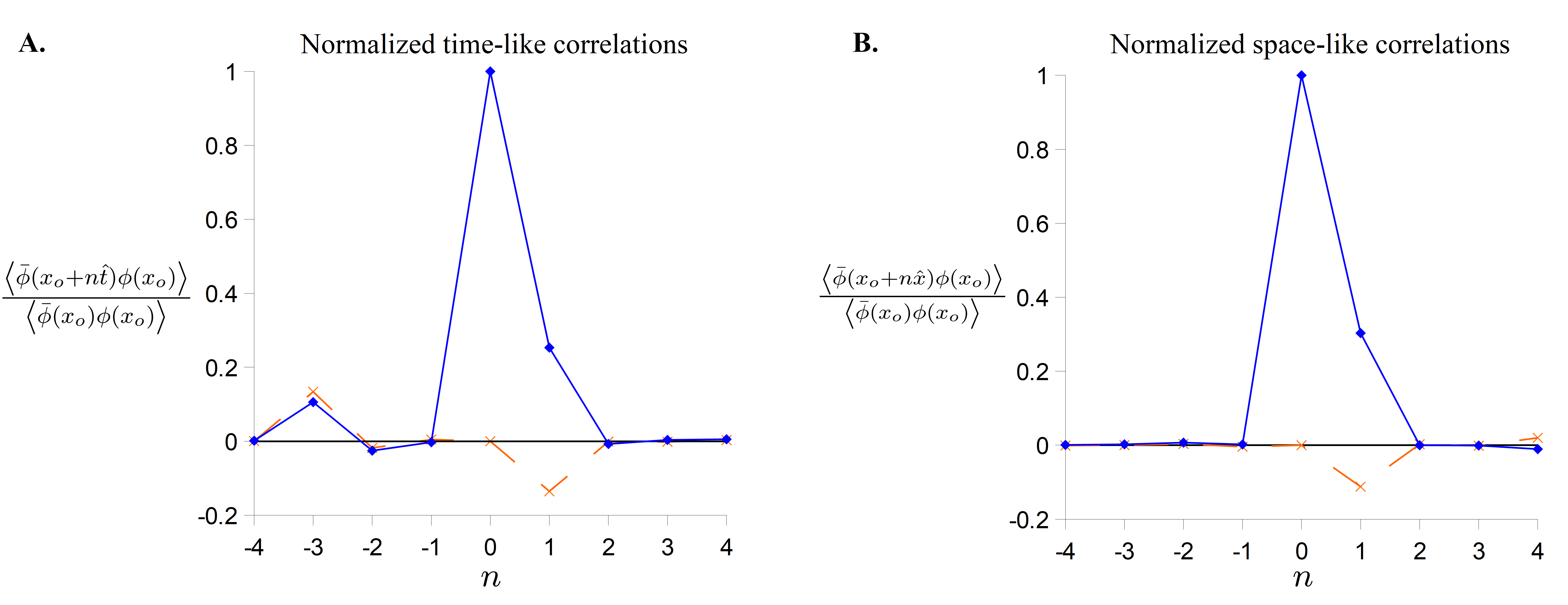}
\caption{Real (\includegraphics[scale=0.1]{Re}) and imaginary (\includegraphics[scale=0.1]{Im}) portions of the time-like (Figure 11A) and space-like (Figure 11B) normalized 2-point correlation function based at the center point ($x_o$) of the Minkowski lattice for the interacting system with $\kappa_1=1$ and $\kappa_2=1$.}
\end{figure*}

Figure 10A depicts the normalized 2-point correlation function for the system with $\kappa_1=0$ and $\kappa_2=1$ evaluated along the time-like line passing through the center point of the Minkowski lattice. Figure 10B depicts the normalized 2-point correlation function for the system with $\kappa_1=0$ and $\kappa_2=1$ evaluated along the space-like line passing through the center point of the Minkowski lattice.

Figure 11A depicts the normalized 2-point correlation function for the system with $\kappa_1=1$ and $\kappa_2=1$ evaluated along the time-like line passing through the center point of the Minkowski lattice. Figure 11B depicts the normalized 2-point correlation function for the system with $\kappa_1=1$ and $\kappa_2=1$ evaluated along the space-like line passing through the center point of the Minkowski lattice.

\begin{figure*}[t!]
\centering
\includegraphics[width=0.95\textwidth]{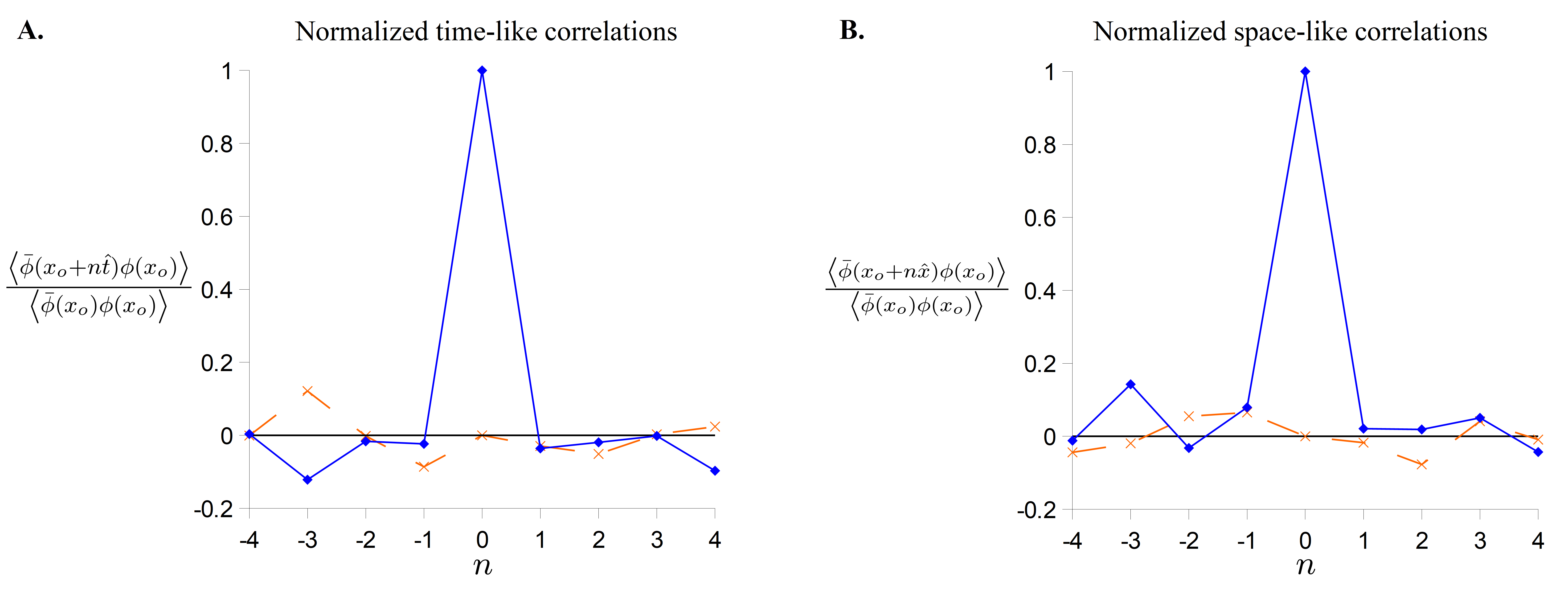}
\caption{Real (\includegraphics[scale=0.1]{Re}) and imaginary (\includegraphics[scale=0.1]{Im}) portions of the time-like (Figure 12A) and space-like (Figure 12B) normalized 2-point correlation function based at the center point ($x_o$) of the Minkowski lattice for the interacting system with $\kappa_1=2$ and $\kappa_2=1$.}
\end{figure*}

Figure 12A depicts the normalized 2-point correlation function for the system with $\kappa_1=2$ and $\kappa_2=1$ evaluated along the time-like line passing through the center point of the Minkowski lattice. Figure 12B depicts the normalized 2-point correlation function for the system with $\kappa_1=2$ and $\kappa_2=1$ evaluated along the space-like line passing through the center point of the Minkowski lattice. Again, the systems of example 3 appear to undergo a phase transition when as the value of $\kappa_1$ increases from 0 to 2 with $\kappa_2$ held constant with a value of 1.

\section*{Discussion}

This article demonstrates that a relativistic statistical field theory can be constructed for fluctuating fields on a discretized Minkowski lattice. The relativistic statistical field theory has an algebra of observables with properties, such as microcausality and the time slice condition, that are similar to the types of properties one would expect of a quantum field theory. Numerical simulations of the fluctuating fields demonstrate proof-of-concept use of this relativistic statistical field theory formulation in non-perturbative calculations, but further development may be required to improve quantitative accuracy. Similarly, further developments may be needed to demonstrate whether a relativistic statistical field theory can be constructed to have all of the properties of quantum field theory and, thus, provide an equivalent reformulation.

\section*{Author Contact Information}
\noindent Brenden.McDearmon@gmail.com
\bibliographystyle{unsrt}
\bibliography{citations}

\end{document}